\documentclass[twocolumn]{aastex63}

\usepackage{amssymb, amsfonts}
\usepackage{amsmath}
\usepackage{booktabs}
\usepackage{tablefootnote}
\usepackage{multirow}
\usepackage{graphicx}
\usepackage{longtable}
\usepackage{commath}
\usepackage{footnote}
\usepackage{tablefootnote}

\newcommand{\msun}{M$_{\odot}$}

\def\h50{\, h_{50}^{-1}}

\def\ltsima{$\; \buildrel < \over \sim \;$}
\def\simlt{\lower.5ex\hbox{\ltsima}}
\def\gtsima{$\; \buildrel > \over \sim \;$}
\def\simgt{\lower.5ex\hbox{\gtsima}}

\defcitealias{chiang15}{Ch15}
\defcitealias{diener15}{D16}
\defcitealias{casey15}{Ca15}
\defcitealias{wang16}{W16}

\received{\today}
\revised{\today}
\accepted{\today}

\submitjournal{ApJ}

\shorttitle{Ly$\alpha$ Emission around a Proto-supercluster at $z \sim 2.47$}
\shortauthors{Huang et al.}

\begin{document}

\title{Evaluating Ly$\alpha$ Emission as a Tracer of the Largest Cosmic Structure at $z \sim 2.47$}

\author{Yun Huang}
\affiliation{Department of Physics and Astronomy, Purdue University, 525 Northwestern Avenue, West Lafayette, IN 47907}
\affiliation{
Visiting astronomer, Kitt Peak National Observatory, National Optical Astronomy Observatory, which is operated by the Association of Universities for Research in Astronomy (AURA) under a cooperative agreement with the National Science Foundation.}
\author{Kyoung-Soo Lee}
\affiliation{Department of Physics and Astronomy, Purdue University, 525 Northwestern Avenue, West Lafayette, IN 47907}
\affiliation{
Visiting astronomer, Kitt Peak National Observatory, National Optical Astronomy Observatory, which is operated by the Association of Universities for Research in Astronomy (AURA) under a cooperative agreement with the National Science Foundation.}
\author{Olga Cucciati}
\affiliation{INAF-Osservatorio di Astrofisica e Scienza dello Spazio di Bologna, Via Piero Gobetti 93/3, I-40129, Bologna, Italy}
\author{Brian Lemaux}
\affiliation{Gemini Observatory, NSF's NOIRLab, 670 N. A`ohoku Place, Hilo, Hawai`i, 96720, USA}
\author{Marcin Sawicki}
\affiliation{Institute for Computational Astrophysics and Department of Astronomy \& Physics, Saint Mary’s University, 923 Robie Street, Halifax, NS B3H 3C3, Canada}
\author{Nicola Malavasi}
\affiliation{Universit\'{e} Paris-Saclay, CNRS, Institut d'Astrophysique Spatiale, 91405, Orsay, France}
\affiliation{
Visiting astronomer, Kitt Peak National Observatory, National Optical Astronomy Observatory, which is operated by the Association of Universities for Research in Astronomy (AURA) under a cooperative agreement with the National Science Foundation.}
\author{Vandana Ramakrishnan}
\affiliation{Department of Physics and Astronomy, Purdue University, 525 Northwestern Avenue, West Lafayette, IN 47907}
\affiliation{
Visiting astronomer, Kitt Peak National Observatory, National Optical Astronomy Observatory, which is operated by the Association of Universities for Research in Astronomy (AURA) under a cooperative agreement with the National Science Foundation.}
\author{Rui Xue}
\affiliation{National Radio Astronomy Observatory, 520 Edgemont Road, Charlottesville, VA 22903, USA}
\affiliation{
Visiting astronomer, Kitt Peak National Observatory, National Optical Astronomy Observatory, which is operated by the Association of Universities for Research in Astronomy (AURA) under a cooperative agreement with the National Science Foundation.}
\author{Letizia P. Cassara}
\affiliation{INAF-IASF Milano, via Alfonso Corti 12, 20129 Milano, Italy}
\author{Yi-Kuan Chiang}
\affiliation{Institute of Astronomy and Astrophysics, Academia Sinica, Taipei 10617, Taiwan}
\author{Arjun Dey}
\affiliation{NSF's National Optical-Infrared Astronomy Research Laboratory, 950 N. Cherry Ave., Tucson, AZ 85719, USA}
\author{Stephen D.J. Gwyn}
\affiliation{Herzberg Astronomy and Astrophysics Research Centre, National Research Council, 5071 W. Saanich Rd. Victoria,
BC, V9E 2E7, Canada}
\author{Nimish Hathi}
\affiliation{Space Telescope Science Institute, Baltimore, MD, USA}
\author{Laura Pentericci}
\affiliation{INAF, Osservatorio Astronomico di Roma, via Frascati 33, I-00078 Monteporzio Catone, Italy}
\author{Moire Prescott}
\affiliation{Department of Astronomy, New Mexico State University, P. O. Box 30001, MSC 4500, Las Cruces, NM, 88003, USA}
\author{Gianni Zamorani}
\affiliation{INAF-Osservatorio di Astrofisica e Scienza dello Spazio di Bologna,
Via Piero Gobetti 93/3, I-40129 Bologna, Italy}

\begin{abstract}

The discovery and spectroscopic confirmation of {\it Hyperion}, a proto-supercluster at $z\sim2.47$, provides an unprecedented opportunity to study distant galaxies in the context of their large-scale environment. We carry out deep narrow-band imaging of a $\approx1^\circ \times1^\circ$ region around {\it Hyperion} and select 157~Ly$\alpha$ emitters (LAEs). The inferred LAE overdensity is $\delta_g\approx40$ within an effective volume of $30\times20\times15$~cMpc$^3$, consistent with the fact that {\it Hyperion} is composed of multiple protoclusters and will evolve into a super-cluster with a total mass of $M_{tot}\approx1.4\times10^{15}$~\msun\ at $z=0$. 
The distribution of LAEs closely mirrors that of known spectroscopic members, tracing the protocluster cores and extended filamentary arms connected to them, suggesting that they trace the same large-scale structure. By cross-correlating the LAE positions with H~{\sc i} tomography data, we find weak evidence that LAEs may be less abundant in the highest H~{\sc i} regions, perhaps because Ly$\alpha$ is suppressed in such regions. The {\it Hyperion} region  hosts a large population of active galactic nuclei (AGN), $\approx$12 times more abundant than that in the field. The prevalence of AGN in protocluster regions hints at the possibility that they may be triggered by physical processes that occur more frequently in dense environments, such as galaxy mergers. Our study demonstrates LAEs as reliable markers of the largest cosmic structures. When combined with ongoing and upcoming imaging and spectroscopic surveys, wide-field narrow-band imaging has the potential to advance our knowledge in the formation and evolution of cosmic structures and of their galaxy inhabitants.

\end{abstract}


\section{Introduction} \label{sec:intro}

% -------------------------------

Galaxy clusters provide useful cosmic laboratories to study how the large-scale environment influences the formation and evolution of galaxies. Existing observations show that cluster galaxies underwent early accelerated formation followed by swift quenching. Since then, they have been evolving passively \citep{stanford98, eisenhardt08, snyder12, martin18}.
In order to elucidate the physical processes regulating galaxy evolution and quenching, we need to rewind the cosmic clock and observe the galaxy growth in young ‘protoclusters’ (i.e., structures that will eventually collapse into galaxy clusters, \citealt{overzier16}). 

Over the years, multiple techniques have been developed to identify protoclusters. These techniques include searching for overdensities of galaxies, e.g., Lyman break galaxies \citep[LBGs, ][]{steidel98,intema06,overzier06,toshikawa16,toshikawa18}, Ly$\alpha$ emitters \citep[LAEs, ][]{matsuda05,kslee14,chiang15,dey16,shi19} and H$\alpha$ emitters \citep[HAEs, ][]{kurk04,hatch11,shimakawa14,cooke14,zheng20}; and rare galaxies as signposts of protoclusters such as radio galaxies \citep{venemans02,venemans07,hayashi12,cooke14}, quasars \citep{wold03,kashikawa07,stevens10,trainor12,hennawi15}, and extended Ly$\alpha$ nebulae \citep[LABs, ][]{yang09,yang10,prescott08,badescu17}. Recently, Planck collaboration identified a large number of protocluster candidates selected as Planck point sources with colors consistent with star-forming  galaxies with cold dust \citep[][]{pccs}. 
These approaches have achieved some success although, in most cases, the sizes of confirmed structures are too small to enable systematic comparison \citep{overzier16, harikane19}. As a result, the relative overlap between samples yielded from various approaches and their selection biases remain poorly understood.

Confirming a protocluster as a galaxy overdensity is challenging due to its large cosmological distances and the expected large spatial extent of protoclusters \citep[e.g., ][]{chiang13, muldrew15}. Indeed, a handful of well-characterized systems subtend up to 10 arcmin across the sky. For this reason, it has been difficult to discern the whole extent of a protocluster and  to appropriately define the galaxy's local environment. As a result, while the number of spectroscopically confirmed protoclusters has been steadily on the rise \citep[see e.g., ][]{overzier16, toshikawa16, toshikawa18, mcconachie22}, the number of systems with reasonable spatial characterization remains low and mostly limited to those in a handful of deep extragalactic fields \citep{hayashino04, dey16}. 

This work is motivated by the discovery and spectroscopic confirmation of {\it Hyperion}, a proto-supercluster at $z\sim 2.47$ \citep{cucciati18} in the COSMOS field. Extensive spectroscopy (e.g., VIMOS Ultra-Deep Survey, or VUDS: \citealt{lefevre15}; zCOSMOS Survey: \citealt{lilly07, lilly09}) and multiwavelength photometry \citep[][]{laigle16,weaver22} have revealed the inner structure of individual galaxy overdensities and the connectivity between them in sufficient detail. The structure is composed of multiple protoclusters and galaxy groups, which were identified by \cite{casey15}, \cite{chiang15}, \cite{diener15} and \cite{wang16} (hereafter, \citetalias{casey15}, \citetalias{chiang15}, \citetalias{diener15}, \citetalias{wang16}, respectively). The discovery of such an  immense cosmic structure and the detailed characterization of its constituents provide an unprecedented opportunity to study the formation of galaxies and protoclusters in the context of the large-scale environment. 

A related question is how galaxies can be used to identify massive cosmic structures and characterize their environment. One good candidate as tracers of the large-scale structure has been LAEs \citep{ouchi20}. 
Existing studies show that LAEs tend to have low stellar masses ($\sim$10$^{8-9}$~\msun), young ages ($\sim$10~Myr), and high ratios of star formation rate to stellar mass \citep[$\sim$10$^{-8}$~yr$^{-1}$, ][]{gawiser06, gronwall07, guaita11, nakajima12}; they are also less dusty than any known galaxy population and hosted by moderate-mass halos \citep{gawiser07, guaita10, kslee14, kusakabe18}. These traits make LAEs the best visible tracers of the underlying matter distribution at high redshift. 
%Several well-known protoclusters mapped with LAEs appear to form features reminiscent of those seen in $N$-body simulations such as cores, groups, filaments, and voids \citep{hayashino04, dey16}. 
If LAEs are good tracers of the large-scale structure, it would follow that large-area LAE surveys may provide an effective pathway to explore the distant universe and learn about the evolution of cluster galaxies at the peak of their formation epoch.

In this work, we select a sample of LAEs fine-tuned to match the redshift span of {\it Hyperion} and study their distribution in and around {\it Hyperion} relative to the environment characterized by other tracers including spectroscopic sources \citep{cucciati18}, photometric redshift selected galaxies \citep{laigle16,weaver22}, H~{\sc i} column density \citep{lee14}, and active galactic nuclei (AGN). 

This paper is organized as follows: in Section~\ref{sec2}, we describe the observation strategy and data reduction. LAE sample selection and validation are described in Section~\ref{sec3}. We construct the LAE spatial distribution and compare it with  various tracers of {\it Hyperion} in Section~\ref{sec4}.
In Section~\ref{sec5}, we estimate the descendant mass of the LAE overdensity and discuss the future directions in the field of cluster formation study. Finally, we summarize the main findings in Section~\ref{sec6}.
Throughout this paper, we adopt a cosmology with $\Omega_M=0.286$, $\Omega_{\Lambda}=0.714$, $h=0.696$, $H_0=100h$ km s$^{-1}$ Mpc$^{-1}$. All magnitudes are in the AB system \citep{oke83}. \\

\section{Observations and Data Reduction} \label{sec2}
 \subsection{Observations and  data reduction} \label{sec:observation}

\begin{figure*}
\epsscale{0.8}
\plotone{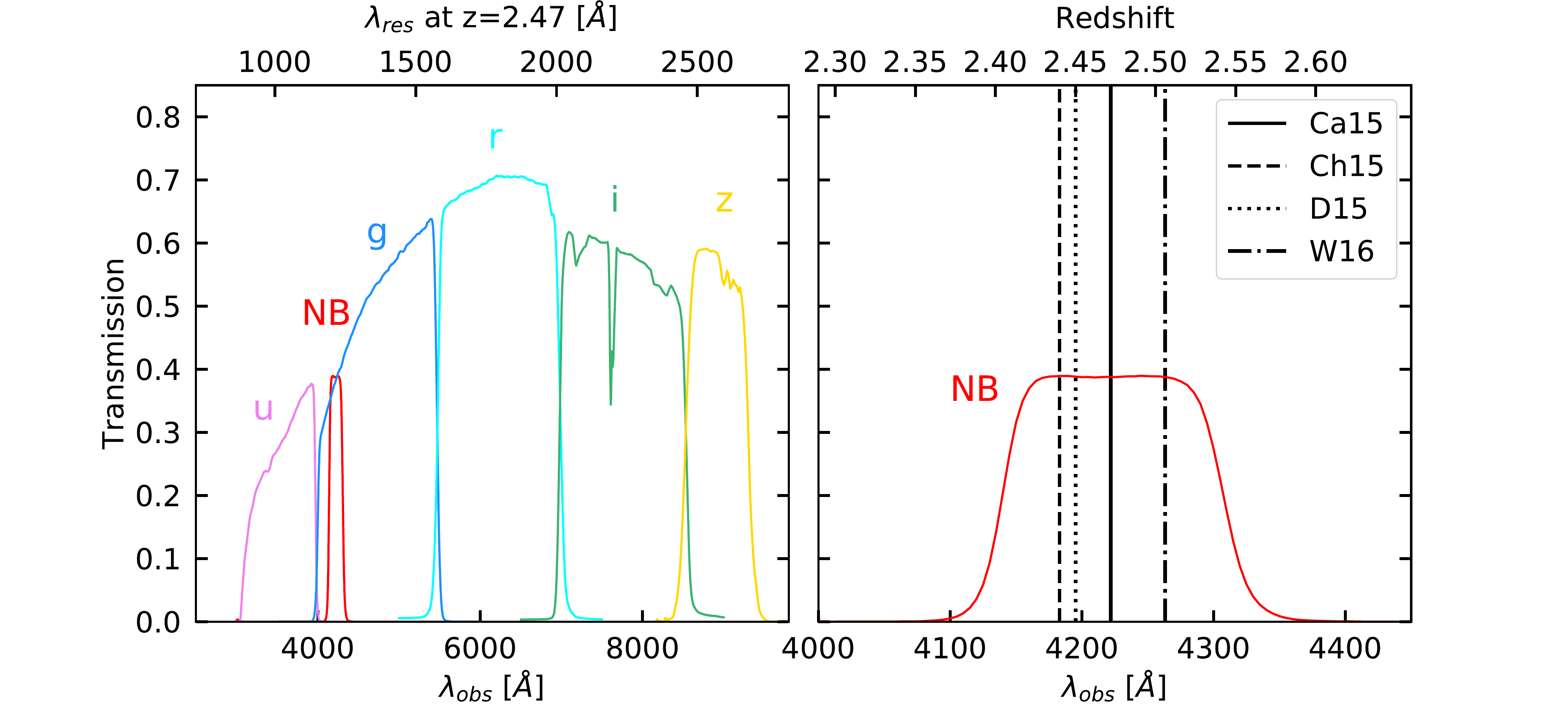}
\caption{
 \textbf{Left:} total throughput data (including filter transmission, mirror, optics, CCD quantum efficiency) of the filters listed in Table~\ref{filters} \textbf{Right:} closeup view of {\it NB422} where Ly$\alpha$ redshift is indicated on top axis. Vertical lines mark the redshifts of the individual protoclusters in the survey field. }
\label{filter}
\end{figure*}

We obtained deep wide-field narrowband images using the One Degree Imager \citep[ODI:][]{odi_old,odi_new} on the WYIN 3.5m telescope\footnote{The WIYN Observatory is a joint facility of the NSF's National Optical-Infrared Astronomy Research Laboratory, Indiana University, the University of Wisconsin-Madison, Pennsylvania State University, the University of Missouri, the University of California-Irvine, and Purdue University.}. ODI has 30 orthogonal transfer array detectors arranged in a $5 \times 6$ configuration, with an intrinsic pixel scale of $0\farcs 11$ pixel$^{-1}$ and a field of view of $40\arcmin \times 48\arcmin$. Our custom narrowband filter (hereafter, {\it NB422}) has a central wavelength of $\lambda_c \approx 4225$~\AA\ and a full-width-at-half-maximum (FWHM) of 170~\AA. The filter is designed to sample redshifted Ly$\alpha$ emission at $z=2.40-2.54$, corresponding to a comoving line-of-sight distance of $\approx$170 cMpc.

The observations were carried out during four separate runs in  2018--2020 using two pointing centers: $\alpha=10^h01^m25.5^s$, $\delta=+02\degr 15\arcmin00\arcsec$ (cosmosE, hereafter) and $\alpha=09^h59^m25.5^s$, $\delta=+02\degr 17\arcmin 24\arcsec$ (cosmosW), J2000. These pointings overlap 0.14~deg in the east-west direction, and together cover a total area of $\sim$0.1 deg$^2$ centered at $\alpha=10^h00^m21.6^s$, $\delta=+02\degr 14\arcmin 24\arcsec$ (cosmosC).
We adopted the 5ODI 9-Point dithering pattern between successive exposures to fill the gaps between CCDs. We discarded the frames with seeing $>1\farcs 3$ and the frames that were taken when the guide star was lost during the exposures. Total exposure times are 19 and 16 hours for the cosmosE and cosmosW pointings, respectively, with the overlapping region receiving the effective exposure of 35 hours. Individual exposures were 10 or 20 minutes long depending on transparency and cloud coverage. 
  
For broad-band imaging data, we utilize the existing  data from the Hyper Suprime-Cam Subaru Strategic Program \citep[HSC-SSP;][]{miyazaki18,komiyama18} second data release \citep{aihara19} and the deep $u$-band data from the CFHT large area U-band deep survey \citep[CLAUDS; ][]{sawicki19}. The broadband data cover the entire survey field imaged with {\it NB422}. Figure \ref{filter} shows the total throughput of all filters, and the filter information is summarized in Table~\ref{filter}. The $5\sigma$ limiting magnitudes reported by HSC-SSP DR2 \citep{aihara19} are for point sources, which are $\sim 0.3$~mag deeper than those measured in a 2\arcsec\ diameter aperture.

\begin{deluxetable}{cccc}
\tablecaption{Filters used in this survey and their characteristics}\label{filters}
\tablehead{
\colhead{Band} & \colhead{Instrument} & \colhead{$m_{5\sigma}$\tablenotemark{a}} & \colhead{FWHM} }
%& & ($5\sigma$, AB) & (\arcsec) }
\startdata
{\it NB422} & WIYN/ODI & 25.4\tablenotemark{b} & 1.0\arcsec \\
$u$  & CFHT/MegaCam & 27.1 & 0.8\arcsec \\
$g$ & Subaru/HSC & 26.9  & 0.7\arcsec \\
$r$ &  Subaru/HSC &  26.6 & 0.9\arcsec \\
$i$ &  Subaru/HSC & 25.3 & 0.7\arcsec \\
\enddata
\tablenotetext{a}{$5\sigma$ limiting magnitude measured in a 2\arcsec\ diameter aperture.}
\tablenotetext{b}{We list the limiting magnitude measured at cosmosW, where the image is shallowest.}
\end{deluxetable}

The raw images of our {\it NB422} data are transferred and processed by the ODI pipeline, Portal and Archive (ODI-PPA; \citealt{gopu14}) where image bias, dark, and pupil ghosts are removed and the images are flat-fielded.

The astrometry is tied to the $Gaia$ second data release \citep{gaia1} using an IRAF task \texttt{msccmatch}.
After correction, the astrometric root-mean-square offsets are $0\farcs1$ for both right ascension and declination. A narrowband image with the best seeing and transparency condition is chosen as the reference image. We then resample each image with a pixel scale of $0\farcs3$ using the tangent point of the reference image using \texttt{SWarp} \citep{swarp}. The scaling factor relative to the reference frame is determined by the IRAF task \texttt{mscimatch}.
Finally, the rescaled reprojected images are combined into a final mosaic using a median combination method. 
Even though median stacking is not an optimal method in maximizing the image depth, it was a necessary choice because the long exposures needed for our observations led to strong amplifier glows in ODI. 
The seeing of the final mosaic image is $1\farcs 0$.

To calibrate the photometric zeropoint of the final mosaic, we use the CLAUDS $u$ and HSC-SSP $g$ band images. To estimate  continuum emission, we calculate the fractional contribution from the flux densities of $u$ and $g$ (denoted as $ug$) bands at the central wavelength of {\it NB422}. The zeropoint of the {\it NB422} is determined by requiring that the median color excess $NB-ug$ is zero for all objects with $0 \leq g-r \leq 1.0$.

\subsection{Photometry} \label{sec:photometry}
  
In order to construct a multi-wavelength photometric catalog,  we first calibrate the astrometry of each broadband image with the $Gaia$ DR2 data. We also match the pixel scale of all images to that of the {\it NB422} image $0\farcs3$~pix$^{-1}$; the broad-band data from the SSP and CLAUDS surveys have the pixel scale of  $0\farcs168$~pix$^{-1}$.
We homogenize the point spread functions (PSFs) of the broadband data with the seeing of {\it NB422} image. A Moffat profile is assumed to fit the PSF of each image with a measured seeing and a fixed parameter $\beta=3$. The convolution kernel is obtained from an IDL routine \texttt{MAX\_ENTROPY}. Each broadband image is then convolved with the respective kernel to create the PSF-matched image.
  
We run \texttt{SExtractor} \citep{sex} on a dual-image mode to create a multi-wavelength catalog where the {\it NB422} image is used for detection ({\tt DETECT\_THRESH}=1.1, {\tt MIN\_AREA}=10). Photometry is performed in all images. Colors are estimated from isophotal flux ({\tt FLUX\_ISO}).  A total of 64,690 sources are detected.

\section{Analysis} \label{sec3}

\subsection{Ly$\alpha$ emitter selection} \label{sec:selection}

In the right panel of Figure~\ref{filter}, we show the {\it NB422} filter transmission. The redshifts of the four largest protoclusters within {\it Hyperion} are indicated as vertical lines, illustrating that any Ly$\alpha$-emitting galaxies that belong to these structures should lie well within the filter transmission. LAEs are selected using the following criteria: 
\begin{equation}\label{eq:selection}
\Sigma_s \geq 2\ \cap \ \rm{S/N}({\it NB422}) \geq 10\ \cap \ {\it NB422}-g \leq -0.5 
\end{equation}
where $\Sigma_s$ is the flux density excess measured relative to the $1\sigma$ photometric scatter expected for zero {\it NB422}$-g$ color. S/N is the signal-to-noise ratio within the isophotal area. The color cut {\it NB422}$-g=-0.5$ corresponds to the rest-frame equivalent width  $W_0=50~$\AA.  
Since our {\it NB422} and $g$-band data are dominated by background noise, we expect both photometric error and $\Sigma_s$ vary  as a function of color:
\begin{equation}\label{eq:lae_select}
m_g-m_{\rm NB}=-2.5\log_{10}(1-\Sigma_s10^{-0.4(ZP-m_{\rm NB})}\sqrt{\sigma_g^2+\sigma_{\rm NB}^2})
\end{equation}
where $ZP$ is the photometric zeropoint of the {\it NB422} image.
Finally, at $z\sim 2.5$, the Lyman limit is redshifted to $\lambda_{\rm obs}\sim 3200$~\AA\ resulting in the flux deficit in the $u$ band. Instead of using the color purely based on the intergalactic H~{\sc i}  absorption, we fine-tune the $u-g$ color criterion based on those of the spectroscopic sources at $z_{\rm spec}$=2.4--2.7 and require: 
\begin{equation}\label{eq:selection2}
u-g \geq 0.9 \times (g-r)-0.2
\end{equation}

\begin{figure}
\epsscale{1.2}
\plotone{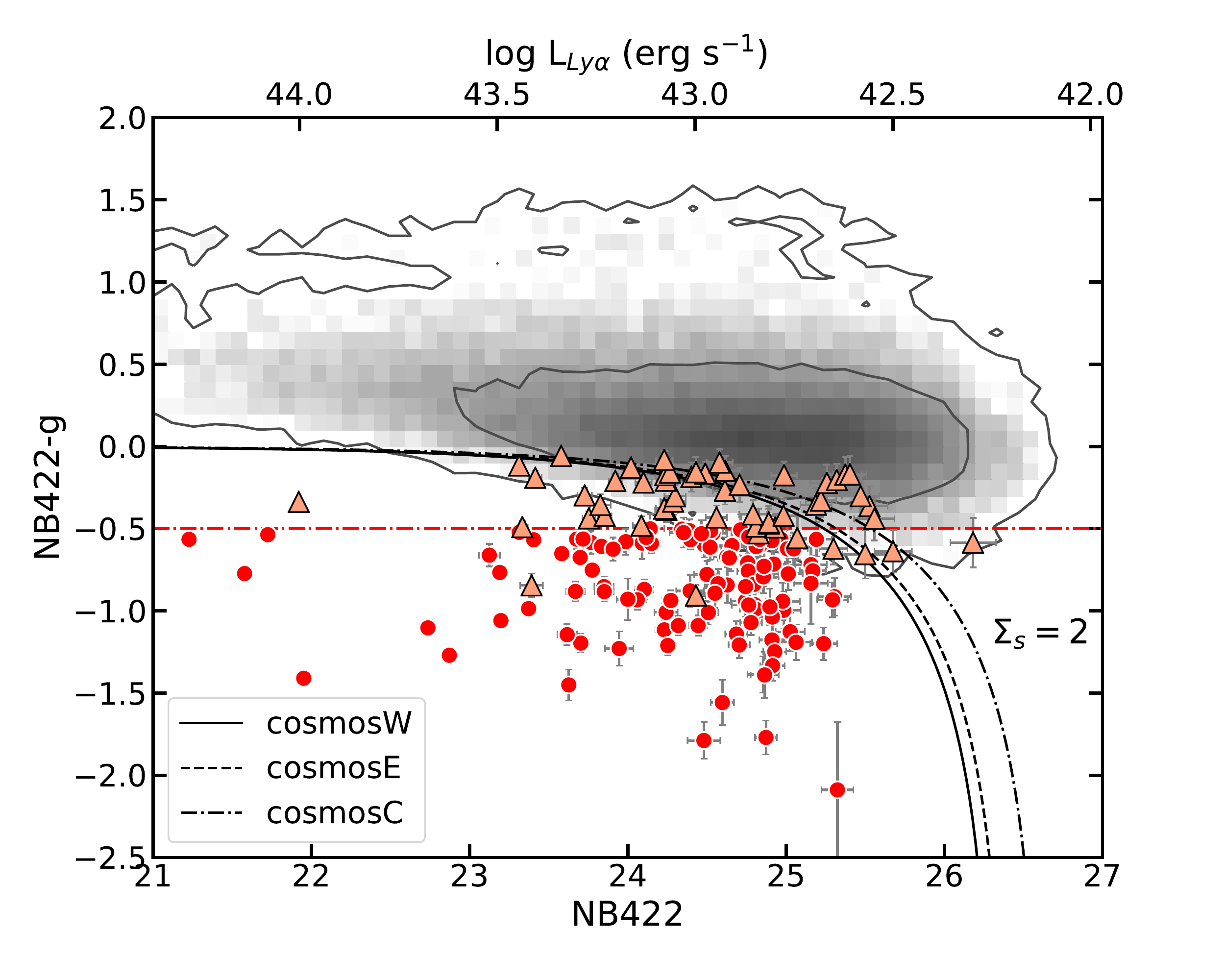}
\caption{
 {\it NB422}$-g$ color as a function of {\it NB422} magnitude. All {\it NB422} sources are represented by the greyscale and the contour lines. Photometric LAE candidates (pLAEs) selected using Equations~\ref{eq:selection} and \ref{eq:selection2} are shown as circles; triangles indicate the $z_{\rm spec}=2.40-2.52$ sources satisfying the same criteria with a modified line excess $\Sigma_s>1$ (sLAEs). The solid, dashed and dash-dotted curves show the $\Sigma_s=2$ lines computed for the cosmosW, cosmosE and cosmosC subfields, respectively. The horizontal dashed line marks the color corresponding to a rest-frame equivalent width $W_0 = 50$~\AA\ at $z=2.45$. 
}
\label{nbmg}
\end{figure}

In Figure~\ref{nbmg}, we show the {\it NB422}$-g$ color vs the narrowband magnitude of all {\it NB}-detected sources. The line-excess criterion applied to each field is shown as black curves\footnote{The sky background noise $\sigma_{\rm NB}$ is estimated separately for the cosmosW ($\alpha<150.02\degr$), cosmosE ($\alpha\geq 150.02\degr$) and the overlapping region, cosmosC. We obtain $\sigma_{\rm NB}=$ 1.08, 1.01 and 0.81~erg~s$^{-1}$~cm$^{-2}$~Hz$^{-1}$ in cosmosW, cosmosE and cosmosC, respectively.}. In principle, our selection criteria (Equations~1--3) could lead to a higher LAE source density in the overlapping region (cosmosC) than that outside it. However, it is clear in Figure~\ref{nbmg} that the stringent color cut (${\it NB422}-g\leq -0.5$) and the S/N requirement primarily drive the LAE selection and that the  fluctuation of $\Sigma_s$ across the field has little impact on the sample selection.  
We select 114 LAEs based on our photometric selection criteria and  refer to them as pLAEs throughout this paper. 

\begin{figure*}
\epsscale{1.15}
\plotone{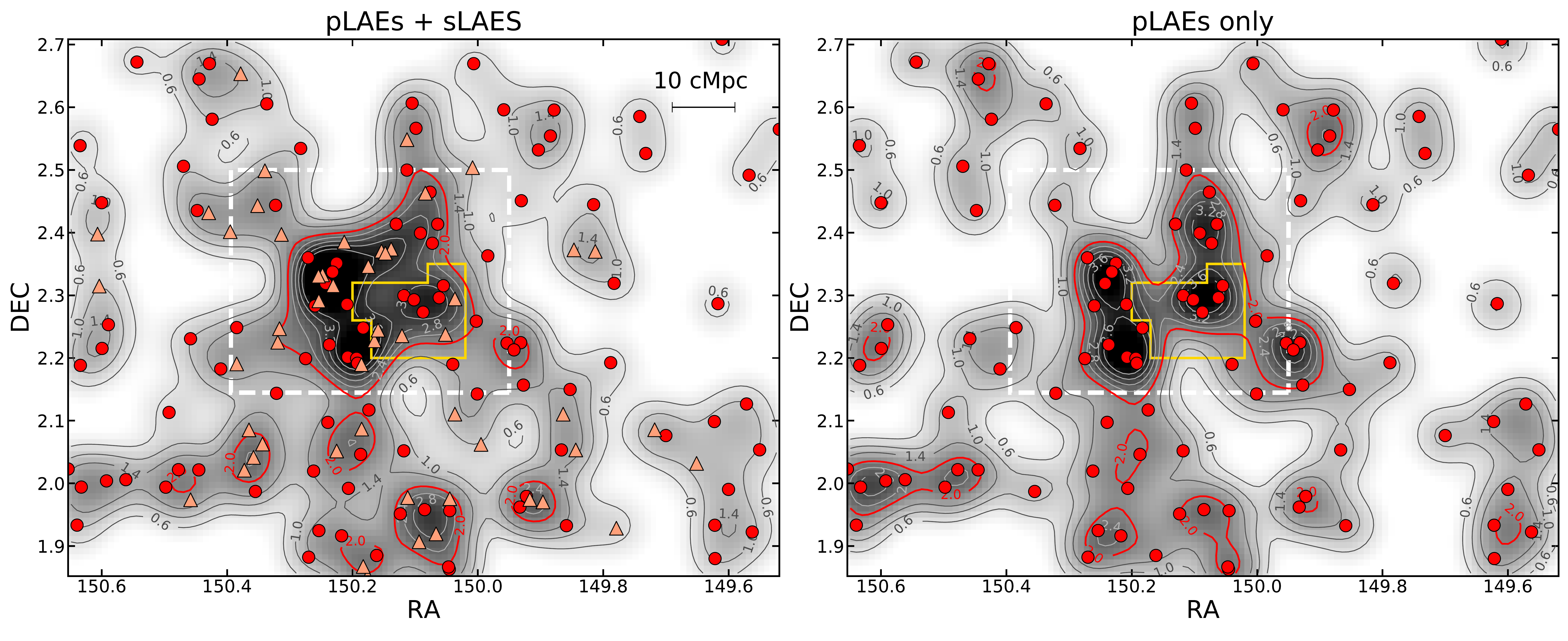}
\caption{
\textbf{Left:} smoothed density map of LAE candidates at $z=2.47$. Red circles and salmon triangles represent pLAEs and sLAEs, respectively. The contours and background grey scale show the surface density levels relative to the field, and the red-highlighted contour ($2.0\bar{\Sigma}$) marks the threshold for protocluster selection. The yellow thick polygon shows the field of view of the HPS survey. The white dashed rectangle shows the field of view of the CLAMATO survey.  \textbf{Right:} smoothed density map constructed by pLAEs only. 
  }
\label{lae_map}
\end{figure*}

Utilizing extensive spectroscopy available in the field, we  select 53 sources at $z_{\rm spec}$=2.40--2.54 requiring a less stringent line excess of $\Sigma_s\geq 1$: i.e., ${\it NB422}-g$ colors are still consistent with having Ly$\alpha$ emission but at a lower equivalent width. We refer to these galaxies as sLAEs. As expected, the majority of the sLAEs have redder ${\it NB422}-g$ colors than the pLAEs (orange triangles in Figure~\ref{nbmg}). A few are too faint and missed by our pLAE selection because they fail to meet either S/N, ${\it NB422}-g$, or $\Sigma_s\geq 2$ requirement. There are a handful of sLAEs that satisfy our selection criteria but are excluded because they are close to bright stars or image boundaries. 
Of the 53 sLAEs, 51 come from the VUDS and zCOSMOS surveys, which cover the region uniformly. The remaining two come from the HETDEX Pilot Survey \citep[HPS:][]{adams11, chiang15} whose coverage is indicated in the right panel of Figure~\ref{lae_map} by a yellow  polygon. 

To summarize, we have selected 114 pLAEs and 53 sLAEs by combining the photometric and spectroscopic catalogs. pLAEs represent higher-EW Ly$\alpha$ emitters selected purely based on photometric criteria while sLAEs are spectroscopic sources at the {\it Hyperion} redshif range with weaker but distinct Ly$\alpha$ emission. Together, our  sample consists of 167 LAEs.

\subsection{Multiwavelength and Spectroscopic data}\label{subsec:multiwavelength}

To validate our sample selection and to quantify the rate of contamination, we compile the existing multi-wavelength and spectroscopic data in the field. For spectroscopy, we merge the redshift catalogs from
the zCOSMOS survey \citep{lilly07,lilly09} and the VIMOS UltraDeep Survey \citep[VUDS:][]{lefevre15}. We will refer to the merged catalog  as the VUDS+zCOSMOS catalog hereafter. The same catalog was  used in \cite{cucciati18}. \\

\noindent {\it Sample Contamination}~~At the central wavelength of our {\it NB422} filter,  the dominant contaminants of the LAE selection are expected to be  [O~{\sc ii}] emitters at $z=0.13$ but the contamination is likely low. At $z=0.13$, the volume covered by our data is negligibly small at $\approx$0.8\% compared to that at $z \sim 2.5$. \citet{ciardullo13} found that at $z<0.2$, [O~{\sc ii}] emitters have a rest-frame equivalent width of $W_{0, \rm{[O~ II]}}= 8 \pm 2$ \AA. Our {\it NB422}$-g$ color cut corresponds to an observed equivalent width of $170$~\AA\ and thus excludes most [O~{\sc ii}] emitters. Assuming that  [O~{\sc ii}]$\lambda$3727 emitters have an equivalent width distribution similar to that measured by \citet{ciardullo13}, we expect $\lesssim 1$ galaxy at $z<0.2$. Indeed, none of our pLAEs are classified as [O~{\sc ii}] emitters in the spectroscopic catalog.

Low-luminosity AGNs with a broad line emission can also contaminate our LAE sample. We cross-match the LAEs with the source lists from the Chandra COSMOS Legacy Survey \citep{civano16, marchesi16} and the XMM-LSS survey \citep{chiappetti05} and find that 24 LAEs have X-ray detections. 
Of these 24, seven with spectroscopic redshifts at $z_{\rm spec}<2.40$ are removed from our LAE sample.
Their blue {\it NB422}$-g$ colors are likely a result of broad C~{\sc iv}, Fe~{\sc ii} and O~{\sc ii} emission lines falling into the {\it NB422} filter \citep{vandenberk01}. 
Of the remaining seventeen, sixteen  lie within the {\it NB422} redshift range and one LAE has no spectroscopic match. The spatial distribution of these seventeen X-ray LAEs is discussed in Section~\ref{subsec:agn}.

We also cross-match the source list with the $Spitzer$ MIPS 24 $\mu$m catalog \citep{sanders07} and the radio data from the 1.4~GHz VLA-COSMOS survey \citep{schinnerer10} and find 4 and 8 matches, respectively. At $z\approx 2.5$, 24 $\mu$m  samples $\lambda_{\rm rest}\approx 7~\mu$m emission of dust heated by starbursts or AGNs \citep{coppin10}. The radio emission is produced by synchrotron radiation from relativistic jets of  supermassive black holes \citep{blandford82,bridle84}. Two MIPS sources and three radio counterparts are spectroscopically confirmed to be within our redshift range. The remaining ones have no spectroscopic detection.

Several LAEs are flagged as AGNs by multiple diagnostics.  LAE\_31446 ($z_{\rm spec}=2.450$) is detected at X-ray, mid-IR and radio wavelengths.  LAE\_8845 (MIPS and radio), LAE\_20628 (X-ray and MIPS), LAE\_3575 and LAE\_34108 (X-ray and radio) also have multiple detections. All lie within the {\it NB422} redshift range except for LAE\_33546 ($z_{\rm spec}=2.146$). 

Regardless of the AGN content, we retain all sources that lie within the {\it NB422} redshift range and also retain six sources with no spectroscopic redshift. After removing seven sources at $z_{\rm spec}<2.40$, 
our catalog contains 160 LAEs (107 pLAEs and 53 sLAEs).\\

\begin{figure}
\epsscale{1.2}
\plotone{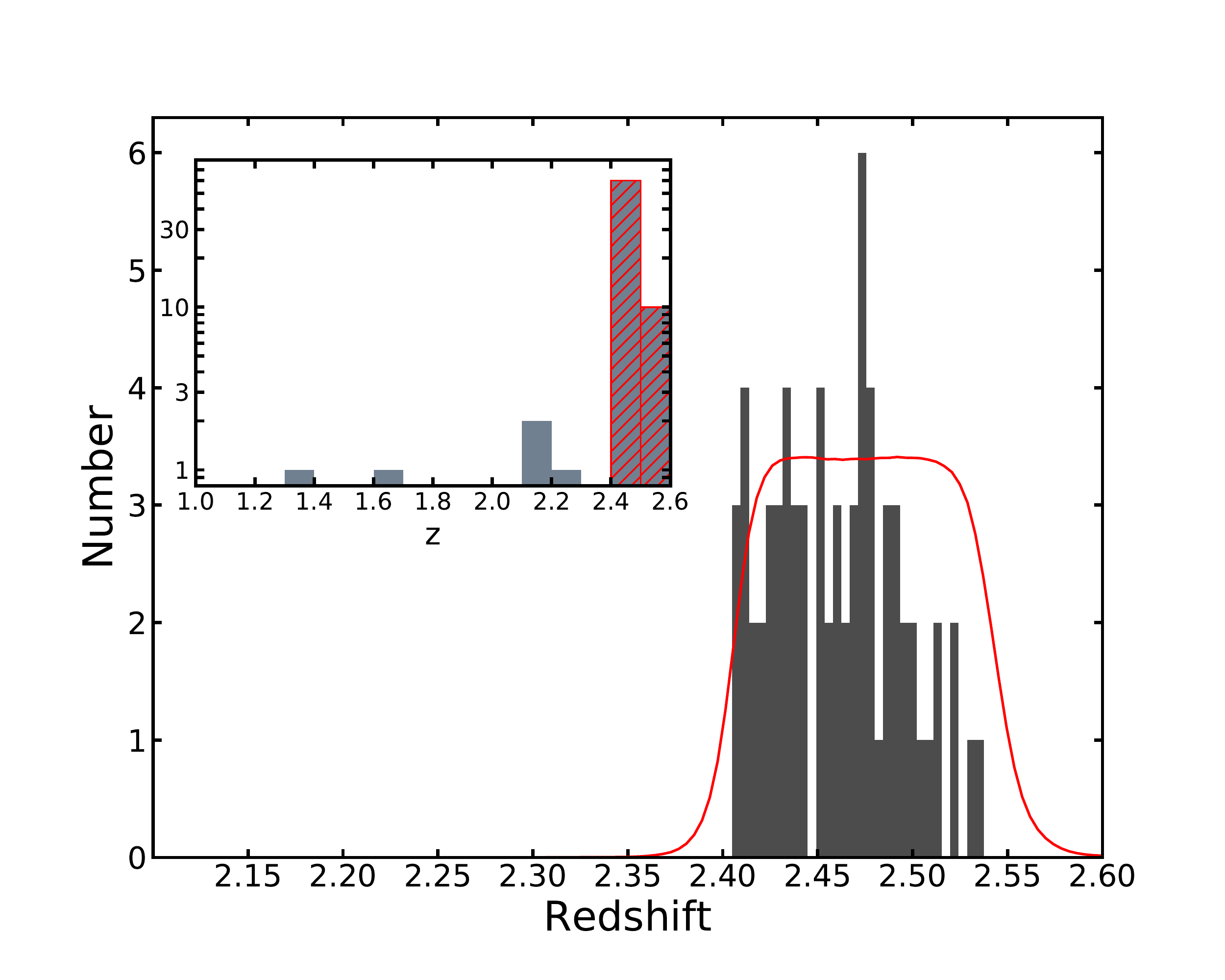}
\caption{
The spectroscopic redshift distribution of the 70 LAEs with spectroscopic counterparts in the redshift range is shown. The red curve shows the {\it NB422} filter transmission with arbitrary normalization. The inset shows the zoomed-out  distribution including 5 low-$z$ redshift interlopers. The red-hatched bars highlight those at {\it Hyperion} redshift.
}
\label{spec}
\end{figure}

\noindent {\it Redshift Distribution}~~ 
Using the VUDS, zCOSMOS, and HPS spectroscopic catalogs, we  validate our LAE selection. 
We only use sources with reliability $\gtrsim 75$\% according to their redshift flag\footnote{The redshift reliability flag equal to X2, X3, X4 or X9. $X=0$ is for galaxies; $X=1$ is for broad-line AGNs; $X=2$ and $X=3$ are the cases where secondary objects fall in the slit and are separable/not separable from the main target \citep{lefevre15}}; for the HPS sources, we use all sources in the COSMOS field because no redshift quality flag is available \citep{adams11}. In Figure~\ref{spec}, we show the redshift histogram of the LAEs with spectroscopic counterparts. The main panel shows the redshift distribution of our LAEs after removing low-$z$ interlopers while the inset shows a broader distribution including them. Nearly half (75) of our LAEs have spectroscopic redshifts. Of these, 70 lie at $z_{\rm spec}$=2.40--2.54, consistent with that expected from the filter transmission curve (red curve). Among the five lower-$z$ interlopers, three are broad-line AGNs identified by VUDS, which we remove from our LAE catalog.

We estimate the purity of our photometric pLAE sample, defined as the fraction of selected sources that fall within the expected redshift range regardless of their spectral types.
Of the 27  pLAEs,  22 are confirmed at $z_{\rm spec}$=2.40--2.54 and 5 at $z_{\rm spec}<2.4$. Thus, the purity of the pLAE sample is 77\%, comparable to the value found by other LAE surveys  at $z=2-6$ \citep[e.g., ][]{ouchi08, nakajima12}.
Our final catalog contains 157 LAEs, which include 104 pLAEs and 53 sLAEs. \\

\section{Large-scale Structure Traced by LAEs} \label{sec4}

\subsection{The sky distribution of LAEs}\label{subsec:density_map}
Our sample consists of 157 LAEs  over the survey  area of 3,455 arcmin$^2$, yielding the mean surface density of $\bar{\Sigma}=(4.5\pm  0.4)\times 10^{-2}$~arcmin$^{-2}$ where the uncertainty reflects the Poisson noise. Using this number as the baseline, we quantify the distribution of local over- and under-densities traced by the LAEs. The 2D distribution is constructed using a PYTHON function \texttt{hist2D}  then smoothed with a Gaussian kernel of a FWHM 10~cMpc. The kernel size is chosen to maximize the density contrast of a protocluster
and is comparable to those adopted in the literature for LAE overdensity identification \citep{badescu17,shi19}. The LAE density map is shown in the  left panel of Figure~\ref{lae_map} where both contours and grayscale  represent the local surface density relative to the field average. The positions of the pLAEs and sLAEs are shown as circles and triangles.

A significant LAE overdensity lies at the center of the field. Taking the $2.0\bar{\Sigma}$ isodensity contour as the boundary, the area of the  overdensity is 226.4~arcmin$^2$ 
%(753.2~cMpc$^2$) 
in which 38 LAEs are enclosed. The number density of LAEs in the region is $0.17\pm 0.03$~arcmin$^{-2}$, a factor of $3.7 \pm 0.5$ times the field average (the overdensity, calculated as $\delta_{\Sigma}\equiv (\Sigma/\bar{\Sigma}-1)$, is $2.7\pm0.5$). 
The likelihood of such an overdensity arising from Poisson fluctuation is $\sim 10^{-11}$. Since the galaxy surface density, $\bar{\Sigma}$,  is strongly affected by the existence of the overdensity, we recompute both field LAE density and the central overdensity after excluding the overdensity region, which are $\bar{\Sigma}=(3.7\pm 0.3)\times 10^{-2}$~arcmin$^{-2}$ and $\delta_\Sigma = 3.6\pm 0.7$, respectively.

Over our survey field, a large fraction of our LAEs have known redshifts from various spectroscopic surveys. Indeed, the number of sLAEs is not negligible compared to that of pLAEs within the central overdensity. To demonstrate that the presence of the overdensity is robust regardless of the inclusion of the spectroscopic sources, we repeat the same procedure but this time only using pLAEs, i.e., the sources which satisfy our LAE criteria (Equation~\ref{eq:selection} and~\ref{eq:selection2}). Our result, shown in the right panel of Figure~\ref{lae_map}, suggests that the same general region stands out as a pLAE overdensity. Similar to previously,  the $2.0\bar{\Sigma}$ iso-density contour has the effective area of 228.4~arcmin$^{2}$ within which 27 pLAEs are found. The central overdensity is $\delta_{\Sigma}=3.8\pm 0.8$ similar to our estimate made including both pLAEs and sLAEs. Both pLAEs and sLAEs broadly trace the same cosmic structure.

\subsection{LAE overdensity vs Hyperion}\label{subsec:hyperion}

\begin{figure*}
\epsscale{1.25}
\centering
\plotone{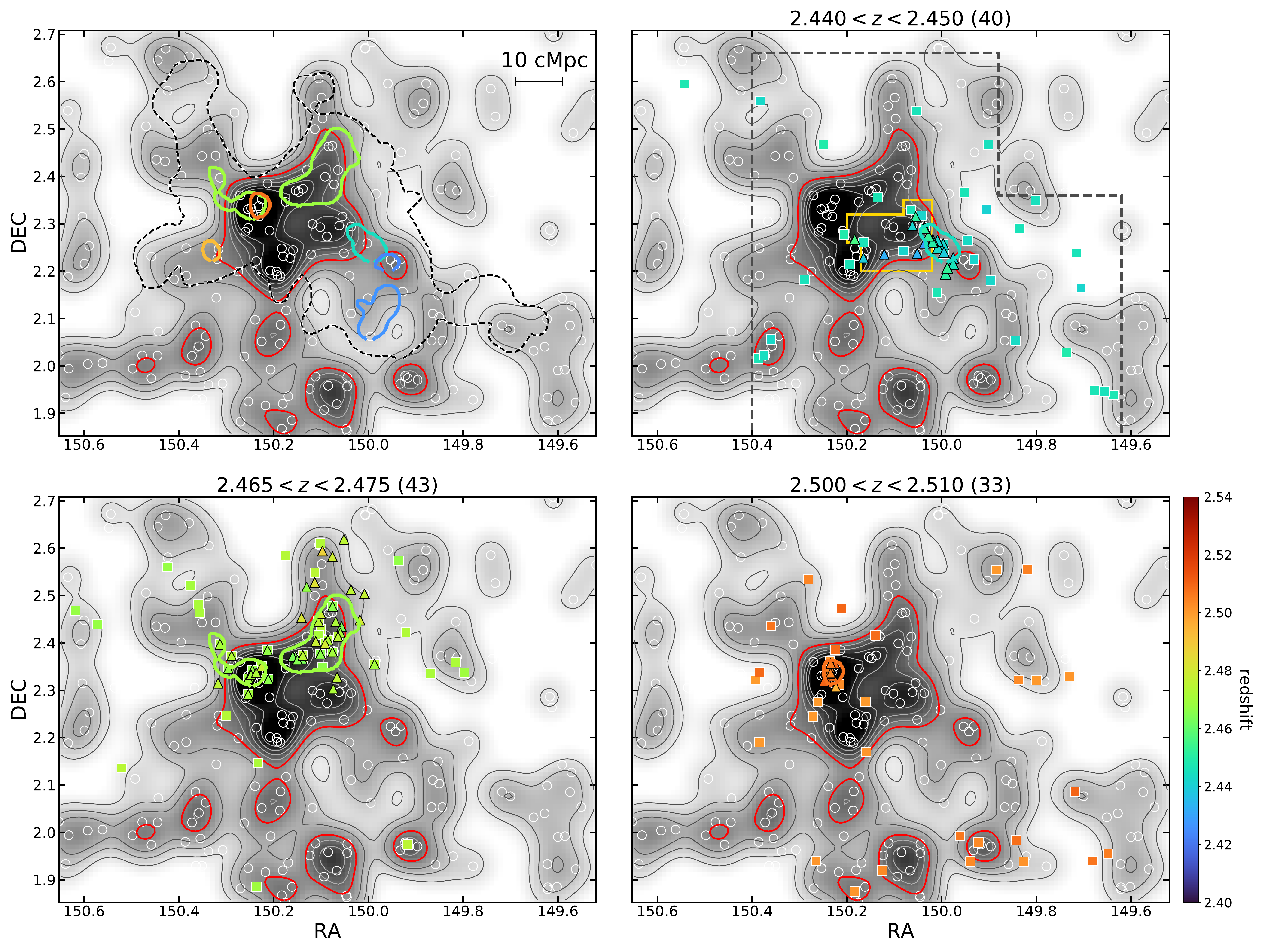}
\caption{
 The sky distribution of all LAEs (white open circles) in the context of {\it Hyperion}. The extent of {\it Hyperion} reported by \citet{cucciati18} is shown in the top left panel as a black dashed line with the known spectroscopic members marked as triangles. The color contours represent the seven density peaks of {\it Hyperion}  color-coded by their redshifts, of which the four largest protocluster members are those reported by \citetalias{chiang15} and \citetalias{diener15} (top right), \citetalias{casey15} (bottom left) and \citetalias{wang16} (bottom right). The redshift range of these individual protoclusters and their number of spectroscopic sources are found at the top of each panel; these galaxies are plotted as squares. The smoothed LAE density maps in the background are identical to that shown in the top left panel of Figure~\ref{lae_map}. The region covered by VUDS and HPS are marked in  the top right panel as dark gray and gold dashed polygons, respectively. All spectroscopic sources are color-coded by redshift as indicated by the color bar. 
  }
\label{laespec_map}
\end{figure*}

In this section, we consider the LAE distribution in the context of the large-scale environment in and around {\it Hyperion} traced by (primarily) more massive star-forming galaxies. Even though we do not know the true underlying matter distribution, spectroscopic identification of galaxies in principle should allow a more inclusive selection in a densely sampled spectroscopic field such as COSMOS. We begin by comparing our LAEs with the density field of {\it Hyperion} measured by \citet{cucciati18} then subsequently with those of the individual protoclusters identified therein \citep{diener15,casey15,chiang15,wang16}. In both cases, spectroscopy largely comes from the VUDS and zCOSMOS surveys. The VUDS survey area is indicated by a gray dashed polygon in  the top right panel of Figure~\ref{laespec_map} while the zCOSMOS coverage is uniform within our survey field.\\

The density field of {\it Hyperion} is estimated by combining the spectroscopic members and photo-z member candidates and performing the two-dimensional Voronoi tessellation on their positions, resulting in a measure of the local surface overdensity, $\delta_{gal, C18}$. We refer  interested readers to \citet{cucciati18} for detail. The extent of {\it Hyperion} is then defined as a contiguous region that rises above $2\sigma_\delta$ where $\sigma_\delta$ is the standard deviation of  log($1+\delta_{gal, C18}$), which is assumed to be a normal distribution. Additionally, they identified seven individual structures whose density peaks rise above $5\sigma_{\delta}$ from the mean density. In the top left panel of Figure~\ref{laespec_map}, we show the projected bounds of {\it Hyperion} (black dashed line) and its seven peaks (solid color contours). Each line corresponding to a peak is color-coded by redshift indicated by the color bar.

The extent of the LAE overdensity (indicated by red contours in Figure~\ref{laespec_map}) is nearly entirely enclosed by the much larger bounds of {\it Hyperion}. Both distributions also feature similar `filamentary arm'-like features stretching out from the main body, two northward and the third in the northeast-southwest direction. These features will be discussed later. Four of the seven {\it Hyperion} peaks reside inside or significantly overlap with the LAE overdensity contour (cyan, green, and darker orange contours in the top left panel of Figure~\ref{laespec_map}). The lowest redshift peaks (blue contours) do not have LAE counterparts possibly because their redshift ranges lie near the blue cutoff of the {\it NB422} transmission (Figures~\ref{filter} and \ref{spec}).

The four structures lying inside the LAE overdensity largely overlap with individual protoclusters reported by \citetalias{chiang15} and \citetalias{diener15} ($z$=2.44--2.45),  \citetalias{casey15}  ($z$=2.47), and  \citetalias{wang16}  ($z$=2.50). In  Figure~\ref{laespec_map} (the top right panel and the bottom panels), we  plot the spectroscopic sources within the redshift range\footnote{Since \citet{cucciati18} used 2D Voronoi tessellation in successive redshift bins, the redshift range we adopt here is unlikely to be identical to that used in their analyses.} of each protocluster to discern their angular extent individually. Galaxy members published in these discovery papers are indicated as triangles while all other spectroscopic sources are shown as squares, both color-coded by redshift. The corresponding density peak identified by \citet{cucciati18} is shown as a color contour. In what follows, we utilize these figures as a visual guide and discuss the angular distribution of our LAEs in the context of {\it Hyperion} protoclusters.\\

\noindent{\it LAEs in \citetalias{chiang15} and \citetalias{diener15} structures ($z$=2.440--2.450)}

Figure~\ref{laespec_map} (top right) shows that the main structure whose center position is $(\alpha, \delta, z) = $ (149.50$^\circ$, 2.25$^\circ$, 2.444). It is elongated stretching in the northeast-southwest direction with a smaller group of galaxies  at $\approx$10~cMpc east of it. This configuration is closely mirrored by the LAEs which extend  further in the southwestern direction and form a compact group on the eastern end of the overdensity. 
The structure coincides with two protoclusters identified by \citetalias{chiang15} and \citetalias{diener15}. The two protoclusters are close in both redshift and angular space, and thus are likely a single broad structure.

The protocluster at $z=2.44$ was first reported by \citetalias{chiang15} as a concentration of nine LAEs from the HPS survey while the $z=2.45$  structure  was initially identified by the zCOSMOS survey as one of the 42 galaxy proto-groups at $z=1.8-3.0$; based on the VLT/FORS2 spectroscopy, \citetalias{diener15} reported eleven spectroscopic member galaxies.

Six of the nine \citetalias{chiang15} sources are recovered by our LAE selection.  Of the remaining three, two have ${\it NB422}-g>0.2$ colors and the third has no {\it NB422} detection. Given the large size of the fibers (4\farcs2), these sources may have been mismatched to other  galaxies. Alternatively, they may be serendipitous detection of fainter galaxies. None of the \citetalias{diener15} sources are LAEs; they are  bright at near-infrared wavelength ($K_s < 24.0$), suggesting that they represent more evolved, more massive galaxies than our LAEs. \\

\noindent{\it LAEs in the \citetalias{casey15} structure ($z$=2.465--2.475)}

Two {\it  Hyperion} density peaks lie at $z\approx 2.47$ with the center positions $(\alpha, \delta, z) = $ (150.09$^\circ$, 2.40$^\circ$, 2.468) and (150.26$^\circ$, 2.34$^\circ$, 2.469). The structure was initially discovered by \citetalias{casey15} as
an overdensity of dusty star-forming galaxies (DSFGs) at $z=2.47$. Extensive follow-up spectroscopy revealed 41 galaxy members including 33 LBGs, 7 DSFGs, and one quasar. Only 9 of these members are recovered by our LAE selection, consistent with the fact that dusty or UV-luminous galaxies tend not to be strong line emitters.

Similar to the \citetalias{diener15} structure, the member galaxies form a linear structure stretching out in the north-south direction. However, the full spectroscopy of the region revealed another long array of galaxies in the east, nearly parallel to the main body. The two are joined together at  the northern end of the LAE overdensity forming a letter `U'.  As such, it appears that the $z\approx 2.47$ structure possibly traces two large overdensities each with a filamentary arm. 
Once again, the LAEs trace the spectroscopic members remarkably well; the contour lines clearly mark the locations of the two filaments stretching northward and the two main overdensities of the $z\approx 2.47$ structure lie well inside the LAE overdensity.  \\

\noindent{\it LAEs in the \citetalias{wang16} structure ($z$=2.500--2.510)}

One of the most compact peaks of {\it Hyperion} is located at $(\alpha, \delta, z) = $ (150.23$^\circ$, 2.34$^\circ$, 2.507). 
\citetalias{wang16} initially reported this structure  as a highly significant ($11.6\sigma$) overdensity of distant red galaxies. Many are populated in a compact region  $\approx$10\arcsec\ in diameter ($\approx$80~kpc physical). The region shows weak extended X-ray emission detected by {\it Chandra}, suggesting a possible presence of a hot proto-intracluster medium.  Seventeen  massive galaxies ($M_{\rm star}\gtrsim 10^{10.5}M_\odot$) are confirmed as spectroscopic members via ALMA and near-IR spectroscopy. Though individually plotted, their locations appear as a single symbol in Figure~\ref{laespec_map} (bottom right) due to their proximity.

Although the LAE overdensity peaks near the highest concentration of DSFG members, it remains unclear how many of the LAEs  belong to $z\approx 2.50$. As can be seen in the top left panel of Figure~\ref{laespec_map}, the region is home to two galaxy overdensities (at $z\approx 2.47$ and $\approx 2.50$). By cross-matching the positions of LAEs with the VUDS/zCOSMOS catalog, we find that two LAEs are found in the outskirts of the structure at $z= 2.53$ while the remainder is at $z=2.47$. Our visual inspection also suggests the least correlation between the spectroscopic sources and the LAEs.

The dearth of LAEs associated with the \citetalias{wang16} structure is in stark contrast with all other protoclusters in the region. Given the tentative X-ray detection and the excess of evolved galaxy populations in the region, it is possible that, in the most massive halo, the formation of low-mass line-emitters is already suppressed (see Section~\ref{subsec:higas} later). However, the relative lack of LAEs in the extended region around the core is more difficult to explain.   \\

\begin{figure*}
\epsscale{1.}
\plotone{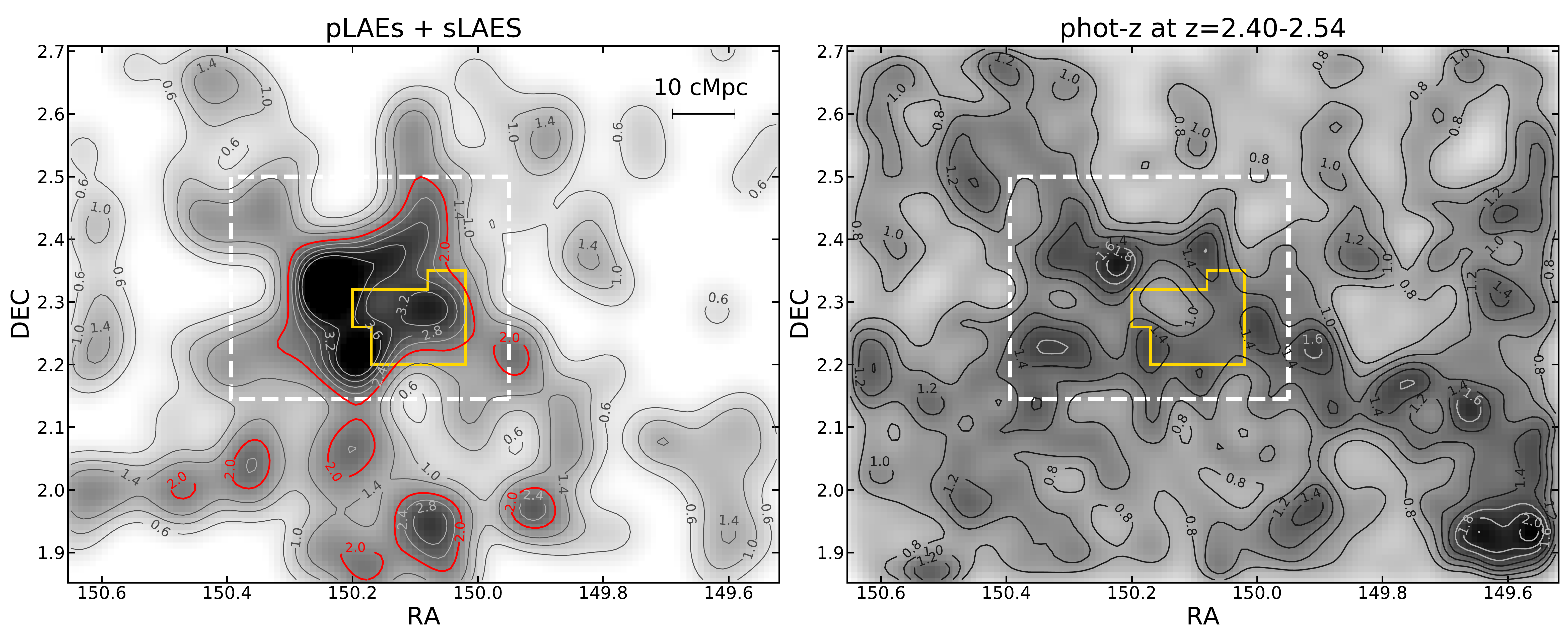}
\caption{
\textbf{Left:} smoothed density map of LAE candidates at $z=2.47$. The contours and background grey scale are identical to the left panel of Figure~\ref{lae_map}. The yellow thick polygon shows the field of view of the HPS survey. The white dashed rectangle shows the field of view of the CLAMATO survey.  \textbf{Right:} smoothed density map using selected photometric galaxies in COSMOS2015 catalog \citep{laigle16}.
  }
\label{phot_map}
\end{figure*}

In addition to the association of LAEs with individual protoclusters, we also perform the Spearman's rank correlation test to determine how well LAEs trace the spec-z sources. To this end, we use the {\tt scipy.stats.spearmanr} routine in PYTHON limiting the sources to lie within  the region $\alpha_{2000}=150.00^\circ-150.45^\circ$ and $\delta_{2000}=2.15^\circ-2.50^\circ$. The redshift range is chosen to be 2.40--2.54 to reflect the redshift selection function defined by the {\it NB422} transmission (Figure~\ref{spec}); only the spectroscopic sources with redshift reliable flag larger than 2 are used. The resultant galaxy samples are then binned to construct a 2D histogram with each cell 16\arcsec\ on a side. 
We obtain the Spearman's coefficients $\rho=0.27$ and $p=8.09\times 10^{-31}$ for the full LAE sample: i.e., the likelihood that the two distributions are correlated ($\rho > 0$) is extremely high at $1-p$.
Since the sLAEs  are selected in part because they lie at this redshift range, we also repeat the test by only using pLAEs and obtain $\rho=0.06$ and $p=0.012$. While the significance is lower, the result suggests that the two distributions are correlated at a 2.5$\sigma$ level. The similarities in the LAEs versus spectroscopic sources strongly hint that the LAEs are excellent tracers of the large-scale structure within massive protoclusters.\\

\subsection{Environment measured by photometric redshifts}\label{subsec:photomap}
Using the public COSMOS2015 catalog \citep{laigle16}, we characterize the large-scale environment traced by photometric galaxies. Although photometric redshift precision is poorer than that achievable by spectroscopy and narrowband imaging, the catalog provides more than half a million galaxies  thereby substantially lowering the Poisson fluctuations of local density measurements. Additionally, the measurements of SFR and stellar masses of each galaxy  can  elucidate the evolutionary stages of galaxies within the structure. Indeed, several studies used the photo-z technique to identify protoclusters \citep{toshikawa12,toshikawa16, chiang14} or to study the density-SFR relation \citep{koyama13b, cooke14} at high redshift. 

The COSMOS2015 sources are selected based on the UltraVISTA DR2 survey $YJHK_S$ bands. The  catalog also includes a suite of deep optical data from CFHT and Subaru and the {\it Spitzer} IRAC 3.6/4.5$\mu$m data from the Spitzer Large Area Survey with Hyper-Suprime-Cam (SPLASH) survey. Although several strips covering an area of 0.6~deg$^2$ were imaged at a greater depth ($K_S=24.7$~AB, 3$\sigma$, 3\arcsec), we limit our analysis to the sources brighter than $K_S=24$ to ensure uniformity across the field, roughly corresponding to $M_{\rm star}\gtrsim 10^{10}M_\odot$ at $z=2.5$.  When compared against the zCOSMOS and VUDS spectroscopic sources at $z>1.5$, we find the redshift uncertainty to be $\sigma_{\Delta z}/(1+z_s) \sim 0.03$ 
and the catastrophic failure ranges between $8$\% and $13$\%, respectively.

To construct the local density map, we treat each galaxy's redshift in a probabilistic manner: i.e.,  the probability density function (PDF) of a given galaxy obeys a normal distribution centered on $z$={\tt zPDF}; the standard deviation is computed as one-half of the distance between the upper and lower 68\% confidence level ({\tt zPDF\_h68} and {\tt zPDF\_l68}). 
We assign a redshift to each galaxy accordingly but only retain  galaxies at $z$=2.40--2.54, typically resulting in  $\approx$2,000 galaxies in the 1.5~deg$^2$ UltraVISTA field. Using these galaxies, the local density map is constructed in an identical manner as in Section~\ref{subsec:density_map}. This procedure is repeated 10,000 times and the final density map is computed as the average over all realizations and is visualized in the right panel of Figure~\ref{phot_map}.

The photo-z density map reasonably matches several features present in the LAE overdensity map including the highest galaxy overdensities within {\it Hyperion} and the voids around it. However, the features reminiscent of `filaments' present in both LAE and spectroscopic distributions are largely missed by the photo-z density map; this is not surprising given that these regions are intermediate in the density range and thus are more likely to be washed away by the large photo-z uncertainties. Additionally, the most significant photo-z overdensity --  
located at the southwestern corner near $(\alpha, \delta)\approx(149.6^\circ, 1.9^\circ)$ -- does not have a counterpart in the LAE overdensity.
The same structure was identified by \citet{chiang14} constructed from an earlier version of the COSMOS photometric redshift catalog. The nature of this galaxy overdensity is  unknown. It could be a protocluster at a slightly lower or higher redshift than {\it Hyperion} such that Ly$\alpha$ emission from its star-forming members falls outside our filter bandpass.

We repeat the same analysis using the COSMOS2020 catalog \citep{weaver22}, an updated version of the COSMOS2015 catalog with an improved photometric redshift precision ($\sigma_{\Delta z}/(1+z_s) \lesssim 0.025$ and the catastrophic failure $\lesssim 8$\%) and find a nearly identical result.

In conclusion, a photometric-redshift technique  offers a  promising avenue of identifying distant protoclusters, but its estimate of local density needs to be taken with caution. Additionally, the regions of interest such as filaments connecting to high-density knots are more clearly traced by line-emitting galaxies.   \\

\subsection{The distribution of LAEs vs H~{\sc i} gas}\label{subsec:higas}

\begin{figure*}
\epsscale{1.}
\centering
\plotone{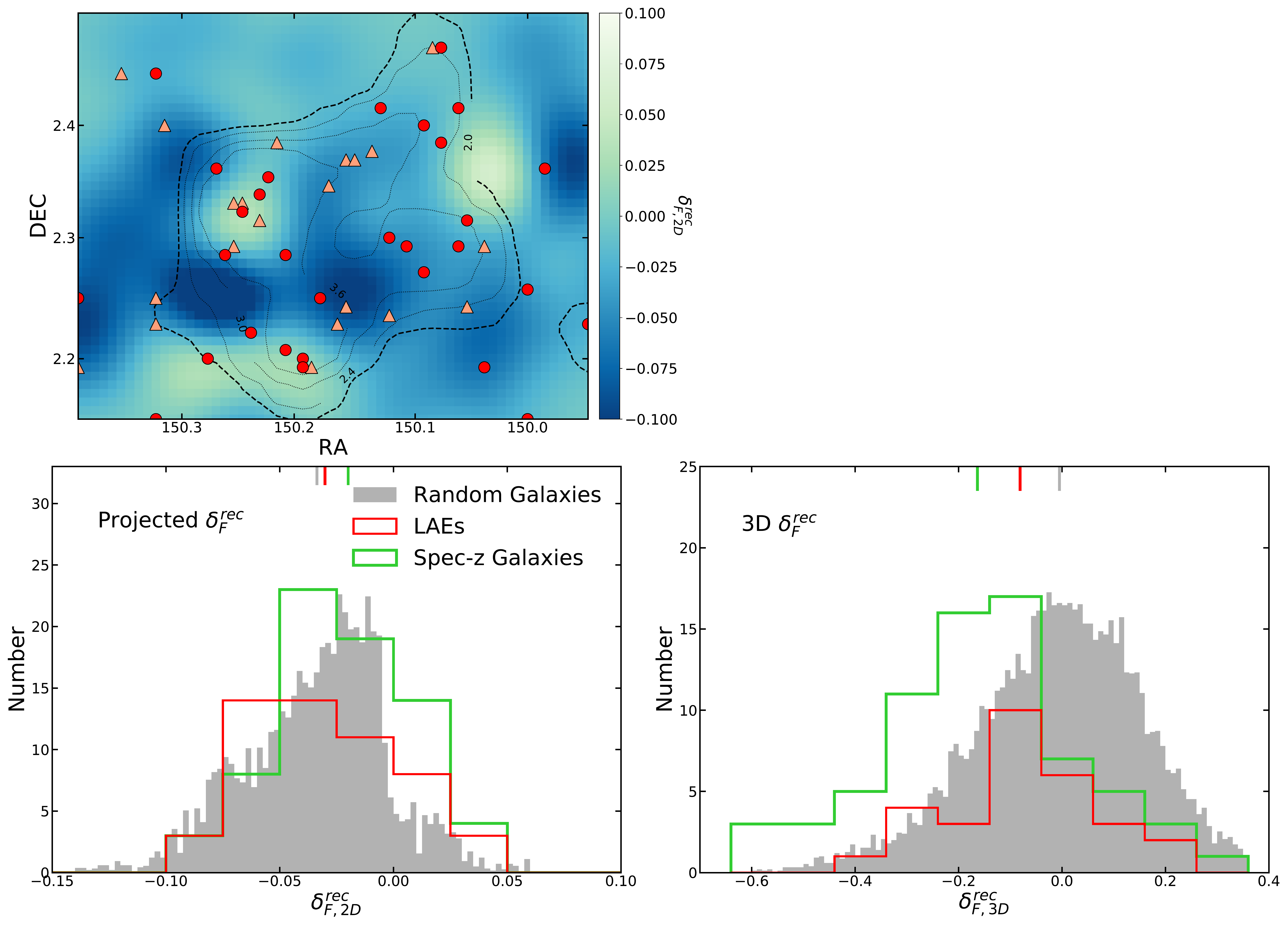}
\caption{
 \textbf{Top left:} the sky distribution of the LAEs within the CLAMATO field of view. Red circles and salmon triangles mark the position of pLAEs and sLAEs, respectively. The contours label the galaxy number density, and the dashed contour highlights the central overdensity. The background color map shows the projected Ly$\alpha$ forest fluctuations $\delta_{F,2D}^{rec}$ over redshift $z=2.40-2.54$. \textbf{Bottom left:}
 histograms of $\delta_{F,2D}^{rec}$ of all LAEs (red steps), spectroscopic members of the {\it Hyperion} (green steps) and random galaxies (gray bars). The mean $\delta_{F,2D}^{rec}$ value of each sample is shown at the top axis. \textbf{Bottom  right:} histograms of $\delta_{F,3D}^{rec}$ of the three galaxy samples. The mean $\delta_{F,3D}^{rec}$ value of each sample is marked at the top axis.
 }
\label{clamato}
\end{figure*}

We compare the LAE distribution with the H~{\sc i} gas distribution as measured by the COSMOS Lyman-Alpha Mapping and Tomography Observations survey \citep[CLAMATO:][]{lee14}. CLAMATO measured Ly$\alpha$ forest absorption in the spectra of  background quasars and star-forming galaxies  at $z=2.3-2.8$   in the region surrounding {\it Hyperion} (see Figure~\ref{lae_map}). The Ly$\alpha$ forest fluctuation is defined as $\delta_F\equiv F/\langle F_z \rangle-1$ where $F$ is the Ly$\alpha$  transmission relative to the  continuum and $\langle F_z \rangle$ denotes the mean transmission as a function of redshift from \cite{faucher08}.
CLAMATO  reconstructed the Ly$\alpha$ forest fluctuation, $\delta^{rec}_F$,  calculated from $\delta_F$ corrected by the noise covariance matrix
with the spatial resolution of 5~cMpc at $z$=2.05--2.55.

To determine the Ly$\alpha$ forest fluctuation at the angular positions of LAEs, we project the 3D CLAMATO data to our survey redshift range $z$=2.40--2.54.  We denote $\delta_{F,2D}^{rec}$ as the reprojected Ly$\alpha$ forest fluctuation while $\delta_{F,3D}^{rec}$ refers to the 3D CLAMATO data cube. In both cases, negative $\delta_{F}^{rec}$ values mean stronger-than-average Ly$\alpha$ absorption  (i.e., high hydrogen column density). 

The top left panel of Figure~\ref{clamato} shows the 2D gas density map where the LAE surface density contours are overlaid. In the bottom left panel, we compare the $\delta_{F,2D}^{rec}$ distributions on  the positions of  LAEs (red) and spectroscopic  members of {\it Hyperion} at $z$=2.40--2.52  (green). We also show the rescaled $\delta_{F,2D}^{rec}$ distribution of 10,000 galaxies at random positions within the field as a reference (grey). 

The result shows  similar mean  values:  $\delta_{F,2D}^{rec}=-0.03$, $-0.02$, and $-0.03$ for the LAEs, spec-z galaxies, and random galaxies, respectively. This is likely because the 2D projection washes away the density fluctuations occurring at protocluster scales ($\Delta z\sim 0.01$ or $r\sim 10-15$~cMpc) as reported by \cite{lee16}. {\it NB422} covers a much larger line-of-sight distance of $\approx 170$~cMpc.

We repeat the same analysis but this time using 29 LAEs that have the CLAMATO data coverage and spectroscopic confirmation to examine the 3D distribution. Random points are drawn in the 3D  volume. The bottom right panel of Figure~\ref{clamato} shows that the locations of LAEs and spectroscopic galaxies are skewed towards more negative $\delta_{F,3D}^{rec}$ values ($-0.08$ and $-0.16$, respectively) suggesting that both are preferentially found in H~{\sc i}-overdensity regions.  The random distribution peaks at $\approx 0$  as expected. While  the $\delta_{F,3D}^{rec}$ distribution peaks at similar values for both LAEs and spec-z sources, the  latter is skewed towards higher H~{\sc i} column densities than the former. The Anderson-Darling test results in $p=0.07$, suggesting the distributions of $\delta^{rec}_{F, 3D}$ are statistically different for LAEs and spec-z sources at $\approx 1.8\sigma$ significance level.  Our result suggests that LAEs systematically avoid the regions of the lowest $\delta_{F,2D}^{rec}$ values (see Figure~\ref{clamato}, top left).

Existing observations show a positive correlation between IGM opacity and galaxy density on Mpc scales \citep{adelberger03,lee16,newman20}. Our result appears to suggest to the contrary, that LAEs tend to avoid the regions with the highest hydrogen column densities. Our finding is consistent with the findings of \cite{momose20}, who measured the cross-correlation function (CCF) between galaxy number density and IGM traced by Ly$\alpha$ forest absorption using the CLAMATO data in conjunction with 19 LAEs at $z_{\rm spec}=2.125-2.225$. They reported a `flattening' of the positive LAE-IGM correlation at  scales $\lesssim$ several~cMpc, hinting  that LAEs are not found in the highest gas- and dark matter density. While more observations are  needed to quantify this effect at a higher significance, our result is consistent with the expectation that Ly$\alpha$ transmission via resonant scattering declines precipitously with the gas column density. 

\begin{figure*}
\epsscale{1.0}
\plotone{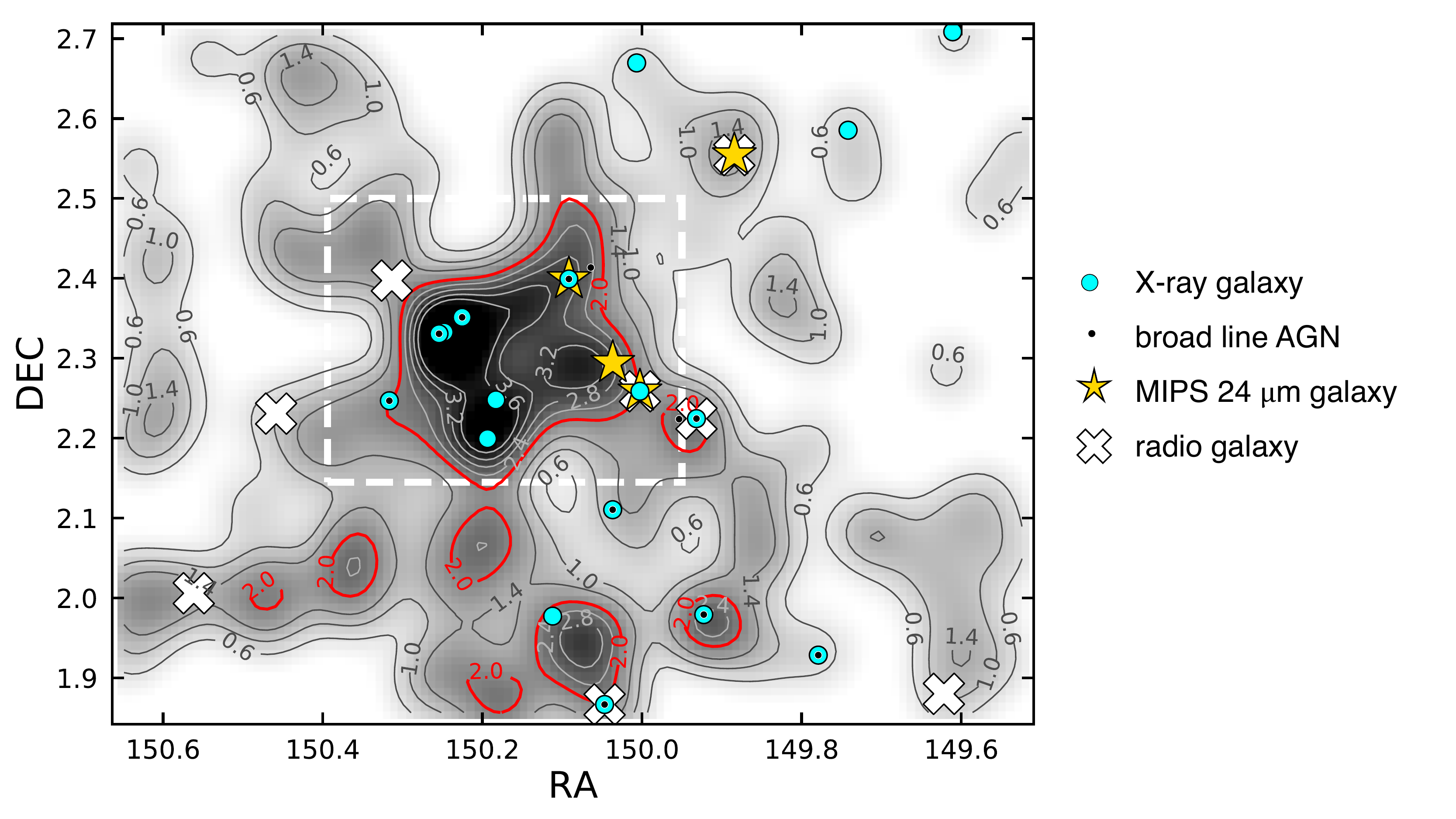}
\caption{
 Sky positions of LAEs with X-ray (cyan circles), MIPS 24 $\mu$m (yellow stars) and VLA (white crosses) detection. The broad line AGNs confirmed by spectroscopic surveys are shown as black dots.
 The smoothed LAE density map in the background and contours are identical to those shown in the top left panel of Figure~\ref{lae_map}. 
  }
\label{agn_map}
\end{figure*}

\subsection{The Prevalence of AGN near the LAE overdensity}\label{subsec:agn}

It is widely accepted that AGN  may play an important role in regulating star formation in galaxies \citep{hopkins14, somerville15, beckmann17, penny18}. At low redshift, existing observations suggest that the level of AGN activity is low in a high-density environment where galaxies tend to be  passively evolving \citep{gisler78, vonderlinden10, mo18}. There is a hint that the AGN--galaxy density relation may reverse at higher redshift \citep{lehmer09, digby10, alberts16, krishnan17} following the reversal of the SFR--density relation   (e.g., \citealt{cooper08, shimakawa18, lemaux20}). 
In this section, we examine the distribution of AGN in relation to the LAE density in and around {\it Hyperion}.

As discussed in Section~\ref{subsec:multiwavelength}, a number of LAEs are classified as AGN based on X-ray, radio, and mid-IR observations. Additionally, existing spectroscopy identified a number of broad-line AGNs at $z=2.40-2.54$. While sources outside the redshift range are removed, we keep those without spectroscopic redshift. In all, we obtain 29 AGNs of which seven do not have spectroscopic confirmation. Figure~\ref{agn_map} shows the sky distribution of these AGNs  overlaid on the LAE surface density map. These AGNs are preferentially  found within the LAE overdensity or along the filamentary structures connected to them traced by LAEs and spectroscopic sources (Section~\ref{subsec:hyperion}).

There is a significant excess  of AGNs in the protocluster region.
Within our survey field, 9 of the 17 X-ray AGNs lie inside the LAE overdensity, yielding the surface overdensity of $\delta_{\Sigma, X}=7.8\pm 3.4$. Similarly, 9 of the 15 broad line AGNs, 3 of the 4 MIPS-detected AGNs and 3 of the 8 radio AGNs reside within the $2.0\bar{\Sigma}$ isodensity contour or near it. When all types of AGNs are combined altogether, the AGN surface overdensity in {\it Hyperion}  rises to $\delta_{\Sigma, all} = 12.3\pm 4.6$. For comparison, the LAE surface overdensity is $\delta_{\Sigma, {\rm LAE}}=3.6\pm 0.7$ (Section~\ref{subsec:density_map}).

In a region identified as a significant overdensity of LAEs, an overdensity of other types of galaxies is  expected. To account for this effect, we calculate the AGN-to-LAE ratio inside and outside the protocluster region and find $0.16\pm 0.07$ and $0.09 \pm 0.03$, respectively, for X-ray-detected AGN. Using the combined AGN sample, the same ratio is $0.37 \pm 0.12$ and $0.13 \pm 0.03$. Though limited by large uncertainties, our result suggests that the level of  AGN excess in {\it Hyperion} is a factor of 2--3 higher than expected.

The enhancement of AGN activity in dense environments has been reported by existing studies at high redshift.  \cite{lehmer09} studied galaxies in the SSA22 protocluster, one of the best studied protocluster system at $z=3.09$, and reported an enhancement of $4.9^{+11.7}_{-3.9}$\% and $7.3^{+17.0}_{-6.2}$\% in the X-ray-detected (log$L_{\rm 8-32 keV}=43.5-44.3$ erg~s$^{-1}$) fraction in LBGs and LAEs, respectively.
A similar trend was reported for lower redshift protoclusters \citep[][]{digby10,krishnan17, tozzi22}.   

The prevalence of AGN in protocluster cores or filaments is consistent with the scenario in which a blackhole activity is triggered due to the major mergers \citep{volonteri03},  which  occur more frequently in dense environment \citep[e.g.,][]{fakhouri09}. Additionally, it is consistent with the fact that AGN are hosted by more massive halos than other galaxies \citep[e.g.,][]{cappelluti12}.
The same scenario is supported by some observations: \citet{hine15} estimated the merger fraction of $0.48\pm 0.10$ in the SSA22 protocluster relative to $0.33\pm 0.08$ in the field. Similarly, \cite{lotz13} found a merger fraction of $0.57^{+0.13}_{-0.14}$ in a $z=1.62$ protocluster, considerably higher than $0.11\pm 0.03$ in the field.
Whether merger event is the dominant mechanism powering AGN activity is under debate. \cite{shah20} found a lack of enhancement in X-ray and IR-detected AGNs among mergers or close pairs while \cite{monson21} reported that LBGs in the SSA22 protocluster exhibit similar merger rates as in the field.

Another phenomenon related to this may be extended Ly$\alpha$ nebulae, or `Ly$\alpha$ blobs'. Several known protoclusters show an excess number of blobs around galaxy overdensities \citep{matsuda04,prescott08,badescu17,shi19}. In SSA22,  an overdensity of both AGN \citep[identified by X-ray and submillimeter observations:][]{umehata19} and Ly$\alpha$ blobs \citep{matsuda04} are found at the `nodes' of filaments traced by LAEs, hinting at the possible link between them. Indeed, AGNs are found in similar locations, close to the outskirts of dense cores \citep{shen21} and filaments \citep{moutard20}. Like AGNs, the clustering strength of blobs indicates that they too are hosted by more massive halos than those hosting normal star-forming galaxies \citep[e.g.,][]{yang10}. Some blobs show direct evidene of being powered by AGN \citep{overzier13, alexander16, momose19} while others point to cold gas accretion in dense environments \citep[e.g.,][]{daddi21,daddi22}. These rare sources provide tantalizing evidence of physical processes uniquely occurring in protocluster environments. Placing robust constraints on the physical connection between protoclusters, AGNs, and blobs will require a much larger scale blind surveys of these objects (see Section~\ref{subsec:outlook}).

\section{Discussion} \label{sec5}

\subsection{Intrinsic galaxy overdensity of {\it Hyperion}}\label{subsec:delta_g}

The {\it NB422} filter has a width 170~\AA\ and  samples $170$~cMpc in the line-of-sight distance at $z\approx 2.4$. This is much greater than the size of a typical protocluster, thus we expect that the spatial overdensity, $\delta_g$, of the structure is greater than the measured surface overdensity, which is diluted by unassociated fore- and background sources. Here, we estimate the true overdensity of the structure traced by our LAEs.

First, we create a  field with a mock protocluster positioned at its center. The `protocluster' region is defined as a rectangle whose area is equal to that enclosed by the $2.0\bar{\Sigma}$ isodensity contour\footnote{The transverse size is $18\times 12$ arcmin$^2$ or $30 \times 20$~cMpc$^2$ at $z=2.4$.} outlined in red in the  left panel of Figure~\ref{lae_map}. In the line-of-sight direction, the observed survey range [2.40, 2.54] is divided into 11 bins with an interval $\Delta z$=0.013,  corresponding to 15~cMpc roughly matching the half-mass diameter of a massive protocluster \citep{chiang13,muldrew15}. In this setup, the true number of galaxies in a protocluster would be $N_{\rm pc}=(1+\delta_g)N_{\rm obs}/(N_{z}+\delta_g)$ where $N_{\rm obs}$ is the total number of LAEs found in the protocluster region; $\delta_g$ is the galaxy overdensity in the redshift space. $N_z=11$ is the number of bins within the observed survey range and $N_{\rm obs}=40$ is the total LAEs detected in the protocluster region. Then we randomly choose a $\delta_g$ value in the range 0--50, compute the number of protocluster- and field galaxies (including their Poisson shot noise), and populate them in the 3D space. The surface overdensity $\delta_{\Sigma}$ is computed from the run, yielding the $\delta_g$--$\delta_\Sigma$ scaling relation given our observational parameters. The procedure is repeated 10,000 times. 

The recovered $\delta_g$--$\delta_\Sigma$ is linear and well-behaved. The LAE surface overdensity  $\delta_\Sigma = 3.7$ implies $\delta_g=39.6^{+5.4}_{-5.1}$, much higher than $\delta_g \sim 3 -10$  expected for a single protocluster \citep[e.g.,][]{lemaux14,cucciati14,topping18,shi19a, hu21}. The unphysically high galaxy overdensity is a direct result of our simplistic assumption that there is a {\it single} protocluster, which is easily disputed by the redshift distribution of our LAEs (Section~\ref{subsec:hyperion}).

We repeat the procedure, but this time, assuming four protoclusters. For the lack of better information about the relative size of the structures within  {\it Hyperion} and their configuration, we assume that each protocluster is confined in a rectangular region 6\arcmin\ on a side  (10.2~cMpc at $z=2.4$) with the identical overdensity. We obtain $\delta_g=12.2^{+2.8}_{-2.0}$. Similarly, assuming seven protoclusters in the region as was found by \citet{cucciati18}, each structure is expected to have $\delta_g=5.9-8.3$. 
While simplistic, our analysis demonstrates that the significance and the angular extent of the LAE overdensity is too great for a single protocluster but instead require at least 3 massive overlapping ones, reinforcing the notion that LAEs provide an effective pathway to find cosmic structures or a complex of structures in distant universe.

Using these $\delta_g$ estimates for each protocluster, we can make a  crude estimate of the total mass locked in them. The main limitation comes from our lack of knowledge in the relative sizes of individual groups and their configuration; additionally, if a substantial fraction of LAEs trace the filaments connecting the main halos, it would lead to an overestimation of the total mass.

Following the method outlined in \cite{steidel98}, the descendant mass of a structure is expressed as:

\begin{equation}\label{mass}
    M_{tot}=\bar{\rho}V_{\rm true}(1+\delta_m)
\end{equation}
where $\bar{\rho}$ is the mean density of the universe, and $V_{\rm true}$ is the true volume that encloses the structure. The matter overdensity, $\delta_m$, is related to galaxy overdensity as $1+b\delta_m=C(1+\delta_g)$ where $C \equiv V_{\rm obs}/V_{\rm true}$ is a correction factor for redshift-space distortion arising from peculiar velocities. In the case of spherical collapse,  $C=1+\Omega_m^{4/7}(z)[1-(1+\delta_m)^{1/3}]$. For $V_{\rm obs}$, we assume the line-of-sight distance 15~cMpc. We adopt a galaxy bias parameter $b=2$ for LAEs \citep{gawiser07, guaita10}. The total mass enclosed therein is $M_{tot}=1.4_{-0.1}^{+0.2} \times 10^{15}$~\msun. The bias value of  $b=2.5$ would decrease the mass by 14\% to $1.2\times 10^{15}$~\msun. Our estimate of the total mass within this volume does not depend on the assumed number of protoclusters.

When compared with existing  measurements, our estimate should be considered with caution. First, it is larger by a factor of $2$ than the sum total of the seven {\it Hyperion} density peaks, $\approx 4.8\times 10^{14}M_\odot$, estimated by \citet{cucciati18}. This is expected because our estimate includes not only the masses of the individual protoclusters but also the filaments and walls between them. Second, our estimate is $\approx$3 times smaller than the total mass of {\it Hyperion}, $\approx 4.8\times 10^{15}M_\odot$, estimated by the same authors. As illustrated in the top left panel of Figure~\ref{laespec_map}, what is considered as the spatial extent of {\it Hyperion}  (black dashed line) is substantially larger than the LAE overdensity in the transverse direction, and possibly in the $z$-direction as well. For the benefit of future protocluster studies, perhaps what is more useful is to test how reliably the present-day mass of a single cosmic structure can be derived through spectroscopy and/or LAE overdensity. To that end, a similar analysis of relatively isolated protoclusters is needed. \\

\subsection{Future outlook: LAEs as beacons of protoclusters}\label{subsec:outlook}

The present study unambiguously demonstrates  LAEs as reliable markers of the largest cosmic structures such as protoclusters and  groups of protoclusters. Further, their distribution can be used to estimate the size, morphology, and present-day mass of the structure and to reveal regions of interest such as dense cores, filaments, and voids. These capabilities will be valuable in any systematic effort to study cosmic structures that lie outside a handful of deep extragalactic fields with extensive spectroscopy. 

The One-hundred-square-degree DECam Imaging in Narrowbands (ODIN) survey has been conducting deep imaging of $\approx$100~deg$^2$ in area using the Dark Energy Camera \citep{decam} to identify LAEs, Ly$\alpha$ nebulae, and forming protoclusters. ODIN targets three cosmic epochs at $z=4.5$, 3.1, and 2.4, straddling the crucial epoch in which the mass assembly in field and cluster galaxies reached its peak \citep[e.g.,][]{madau14,chiang17}.  Over seven fields\footnote{The targeted fields include all four LSST Deep Drilling Fields, two southern Euclid Deep fields, Deep2-3, and HETDEX-SHELA.}, ODIN will sample a total of $\approx$0.25~Gpc$^3$ in cosmic volume in which $\approx 130,000$ LAEs and  $\approx$45 (600) progenitors of Coma- and Virgo-like clusters ($M_{tot}>10^{15}$~\msun and $(3-10)\times 10^{14}$~\msun, respectively) are expected. The details of the survey will be presented in an upcoming paper (K.-S. Lee et al., in preparation).

When combined with the new generation of wide-field imaging and spectroscopic experiments such as LSST, Roman, LMT/TolTEC, HETDEX, and DESI,  ODIN will open up a new window into the formation of cosmic structures and the role of large-scale environment on the galaxy constituents therein. 
First, the physical connection between protoclusters, AGNs, Ly$\alpha$ nebulae, and passive galaxies can be quantified and delineated in a statistically meaningful manner. Measurements of their occurrences, preferred locations, and physical properties can be used to place strong constraints on the physical origin and the timescale of these phenomena.  

Second, a robust sample of protoclusters with accurate redshifts will facilitate comparative studies. Recently, large samples of protocluster candidates have been  identified as LBG overdensities \citep[e.g.,][]{toshikawa16,toshikawa18} and cold dust emission in the far-IR \citep[e.g.,][]{planck_cold_source,planck_cold_source_v2}.  However, the intersection between these samples and the  selection efficiency and bias of different techniques are poorly understood \citep[but see, e.g.,][]{negrello17,gouin22} 

Finally, a large statistical sample of ODIN protoclusters will allow us to measure the average  star formation activity expected in a protocluster and how it scales with their present-day masses over cosmic time. For example, deep wide-field mm surveys such as the LMT/TolTEC Public Legacy survey will not only detect individual high star-formers but also can be used to estimate the total (dust-obscured) SFR per cluster via stacking analysis. Such measurements can resolve or further highlight the  current tension that exists between the observed and predicted level of star formation in protoclusters \citep[e.g.][]{lim21}. \\

\section{Summary} \label{sec6}

In this paper, we carry out a comprehensive study of galaxies  inhabiting {\it Hyperion}, the first spectroscopically confirmed  structure that hosts multiple protoclusters at $z=2.4-2.5$, which will likely evolve into a super cluster of galaxies by the present-day universe. To trace its LSS, we obtain  deep narrow-band observations of a $\approx 1^\circ \times 1^\circ$ of the region using {\it NB422} ($\lambda_c \approx 4225$~\AA, $\Delta\lambda=170$~\AA) and  select 157 LAE candidates at the {\it Hyperion} redshift (Section~\ref{sec:selection}). The new data combined with the extensive spectroscopic and photometric observations available in the field provide a unique opportunity to closely examine how galaxies of different types trace the same large-scale structure. 
Our main findings are listed below: \\

-- A significant excess of LAEs occupies the region of {\it Hyperion} with a surface overdensity $\delta_\Sigma=3.6$ (Section~\ref{subsec:density_map}). This corresponds to the combined spatial galaxy overdensity $\delta_g\approx 40$ within an effective volume of $30\times 20 \times 15$~cMpc$^3$. The level and the extent of the overdensity is too large for a single structure, strongly suggesting that multiple protoclusters must be embedded in {\it Hyperion}. The LAE overdensity will evolve into a super cluster of galaxies with a total mass of $M_{tot} \approx 1.4\times 10^{15}$~\msun\ distributed between the main halos hosting each protocluster and the filaments connecting them (Section~\ref{subsec:delta_g}). \\

-- Taking advantage of the densely sampled spectroscopic data, we evaluate the feasibility of LAEs as a probe of the LSS in and around {\it Hyperion} (Section~\ref{subsec:hyperion}).  The  distributions of LAEs and spectroscopic sources are remarkably similar, suggesting that LAEs are excellent tracers of the LSS (Figure~\ref{laespec_map}).
In particular, the LAE distribution clearly marks several density peaks  and extended structures  in {\it Hyperion}.
However, we also find tentative evidence that LAEs may be poor tracers of the most evolved systems \citep[e.g., see][]{wang16}.\\

-- Using the COSMOS2015 and COSMOS 2020 photometric redshift catalogs, we construct a local density map at  $z=2.40-2.54$. The photo-z map fares reasonably well in finding several strongest features of {\it Hyperion}, including the highest galaxy overdensities and the lowest-density voids around it. However, lower-density features of interest such as filaments may be better traced by line-emitting galaxies (Figure~\ref{phot_map}). \\

-- By cross-correlating the LAE positions with the H~{\sc i} tomography data from the CLAMATO survey \citep{lee16}, we find that LAEs tend to lie in the regions of moderate H~{\sc i}  overdensities while avoiding the highest H~{\sc i} density regions (Figure~\ref{clamato}). Our finding is in line with existing studies \citep[e.g.,][]{momose20} and is consistent with the expectation that the suppression of Ly$\alpha$ transmission may be substantial in the highest-density regions due to resonant scattering. \\

-- In and around {\it Hyperion}, we find a significant overdensity of AGN (Section~\ref{subsec:agn}, Figure~\ref{agn_map}) selected based on X-ray/radio/mid-IR emission or broad emission line features including Ly$\alpha$. When all types of AGN are considered together, the surface overdensity is measured to be $\delta_{\Sigma, all}\approx 12$, i.e., $3.5$ times larger than the LAE surface overdensity of the region, $\delta_{\rm \Sigma, LAE} = 3.6$. The prevalence of AGN in protocluster regions hints at the possibility that they may be triggered by physical processes  that occur more frequently in dense environments, such as galaxy mergers.

\begin{acknowledgments}

The authors acknowledge financial support from NASA through the Astrophysics Data Analysis Program, grant number 80NSSC19K0582.

The observation is based on observations at Kitt Peak National Observatory, NSF's National Optical-Infrared Astronomy Research Laboratory (NOIRLab Prop. ID: 2018A-0446, 2018B-0376, 2019A-0372, 2019B-0483; PI: K.-S. Lee), which is operated by the Association of Universities for Research in Astronomy (AURA) under a cooperative agreement with the National Science Foundation. The $u$-band data were obtained and processed as part of the CFHT Large Area U-band Deep Survey (CLAUDS), which is a collaboration between astronomers from Canada, France, and China described in Sawicki et al. (2019, [MNRAS 489, 5202]).  CLAUDS is based on observations obtained with MegaPrime/ MegaCam, a joint project of CFHT and CEA/DAPNIA, at the CFHT which is operated by the National Research Council (NRC) of Canada, the Institut National des Science de l'Univers of the Centre National de la Recherche Scientifique (CNRS) of France, and the University of Hawaii. CLAUDS uses data obtained in part through the Telescope Access Program (TAP), which has been funded by the National Astronomical Observatories, Chinese Academy of Sciences, and the Special Fund for Astronomy from the Ministry of Finance of China. CLAUDS uses data products from TERAPIX and the Canadian Astronomy Data Centre (CADC) and was carried out using resources from Compute Canada and Canadian Advanced Network For Astrophysical Research (CANFAR).

\end{acknowledgments}

\bibliographystyle{aasjournal}
\bibliography{refs}

\begin{thebibliography}{}
\expandafter\ifx\csname natexlab\endcsname\relax\def\natexlab#1{#1}\fi
\providecommand{\url}[1]{\href{#1}{#1}}
\providecommand{\dodoi}[1]{doi:~\href{http://doi.org/#1}{\nolinkurl{#1}}}
\providecommand{\doeprint}[1]{\href{http://ascl.net/#1}{\nolinkurl{http://ascl.net/#1}}}
\providecommand{\doarXiv}[1]{\href{https://arxiv.org/abs/#1}{\nolinkurl{https://arxiv.org/abs/#1}}}

\bibitem[{{Adams} {et~al.}(2011){Adams}, {Blanc}, {Hill}, {Gebhardt}, {Drory},
  {Hao}, {Bender}, {Byun}, {Ciardullo}, {Cornell}, {Finkelstein}, {Fry},
  {Gawiser}, {Gronwall}, {Hopp}, {Jeong}, {Kelz}, {Kelzenberg}, {Komatsu},
  {MacQueen}, {Murphy}, {Odoms}, {Roth}, {Schneider}, {Tufts}, \&
  {Wilkinson}}]{adams11}
{Adams}, J.~J., {Blanc}, G.~A., {Hill}, G.~J., {et~al.} 2011, \apjs, 192, 5,
  \dodoi{10.1088/0067-0049/192/1/5}

\bibitem[{{Adelberger} {et~al.}(2003){Adelberger}, {Steidel}, {Shapley}, \&
  {Pettini}}]{adelberger03}
{Adelberger}, K.~L., {Steidel}, C.~C., {Shapley}, A.~E., \& {Pettini}, M. 2003,
  \apj, 584, 45, \dodoi{10.1086/345660}

\bibitem[{{Aihara} {et~al.}(2019){Aihara}, {AlSayyad}, {Ando}, {Armstrong},
  {Bosch}, {Egami}, {Furusawa}, {Furusawa}, {Goulding}, {Harikane}, {Hikage},
  {Ho}, {Hsieh}, {Huang}, {Ikeda}, {Imanishi}, {Ito}, {Iwata}, {Jaelani},
  {Kakuma}, {Kawana}, {Kikuta}, {Kobayashi}, {Koike}, {Komiyama}, {Li},
  {Liang}, {Lin}, {Luo}, {Lupton}, {Lust}, {MacArthur}, {Matsuoka}, {Mineo},
  {Miyatake}, {Miyazaki}, {More}, {Murata}, {Namiki}, {Nishizawa}, {Oguri},
  {Okabe}, {Okamoto}, {Okura}, {Ono}, {Onodera}, {Onoue}, {Osato}, {Ouchi},
  {Shibuya}, {Strauss}, {Sugiyama}, {Suto}, {Takada}, {Takagi}, {Takata},
  {Takita}, {Tanaka}, {Terai}, {Toba}, {Uchiyama}, {Utsumi}, {Wang}, {Wang}, \&
  {Yamada}}]{aihara19}
{Aihara}, H., {AlSayyad}, Y., {Ando}, M., {et~al.} 2019, \pasj, 106,
  \dodoi{10.1093/pasj/psz103}

\bibitem[{{Alberts} {et~al.}(2016){Alberts}, {Pope}, {Brodwin}, {Chung},
  {Cybulski}, {Dey}, {Eisenhardt}, {Galametz}, {Gonzalez}, {Jannuzi},
  {Stanford}, {Snyder}, {Stern}, \& {Zeimann}}]{alberts16}
{Alberts}, S., {Pope}, A., {Brodwin}, M., {et~al.} 2016, \apj, 825, 72,
  \dodoi{10.3847/0004-637X/825/1/72}

\bibitem[{{Alexander} {et~al.}(2016){Alexander}, {Simpson}, {Harrison},
  {Mullaney}, {Smail}, {Geach}, {Hickox}, {Hine}, {Karim}, {Kubo}, {Lehmer},
  {Matsuda}, {Rosario}, {Stanley}, {Swinbank}, {Umehata}, \&
  {Yamada}}]{alexander16}
{Alexander}, D.~M., {Simpson}, J.~M., {Harrison}, C.~M., {et~al.} 2016, \mnras,
  461, 2944, \dodoi{10.1093/mnras/stw1509}

\bibitem[{{Beckmann} {et~al.}(2017){Beckmann}, {Devriendt}, {Slyz}, {Peirani},
  {Richardson}, {Dubois}, {Pichon}, {Chisari}, {Kaviraj}, {Laigle}, \&
  {Volonteri}}]{beckmann17}
{Beckmann}, R.~S., {Devriendt}, J., {Slyz}, A., {et~al.} 2017, \mnras, 472,
  949, \dodoi{10.1093/mnras/stx1831}

\bibitem[{{Bertin} \& {Arnouts}(1996)}]{sex}
{Bertin}, E., \& {Arnouts}, S. 1996, \aaps, 117, 393,
  \dodoi{10.1051/aas:1996164}

\bibitem[{{Bertin} {et~al.}(2002){Bertin}, {Mellier}, {Radovich}, {Missonnier},
  {Didelon}, \& {Morin}}]{swarp}
{Bertin}, E., {Mellier}, Y., {Radovich}, M., {et~al.} 2002, Astronomical
  Society of the Pacific Conference Series, Vol. 281, {The TERAPIX Pipeline},
  ed. D.~A. {Bohlender}, D.~{Durand}, \& T.~H. {Handley}, 228

\bibitem[{{Blandford} \& {Payne}(1982)}]{blandford82}
{Blandford}, R.~D., \& {Payne}, D.~G. 1982, \mnras, 199, 883,
  \dodoi{10.1093/mnras/199.4.883}

\bibitem[{{Bridle} \& {Perley}(1984)}]{bridle84}
{Bridle}, A.~H., \& {Perley}, R.~A. 1984, \araa, 22, 319,
  \dodoi{10.1146/annurev.aa.22.090184.001535}

\bibitem[{{B{\u{a}}descu} {et~al.}(2017){B{\u{a}}descu}, {Yang}, {Bertoldi},
  {Zabludoff}, {Karim}, \& {Magnelli}}]{badescu17}
{B{\u{a}}descu}, T., {Yang}, Y., {Bertoldi}, F., {et~al.} 2017, \apj, 845, 172,
  \dodoi{10.3847/1538-4357/aa8220}

\bibitem[{{Cappelluti} {et~al.}(2012){Cappelluti}, {Allevato}, \&
  {Finoguenov}}]{cappelluti12}
{Cappelluti}, N., {Allevato}, V., \& {Finoguenov}, A. 2012, Advances in
  Astronomy, 2012, 853701, \dodoi{10.1155/2012/853701}

\bibitem[{{Casey} {et~al.}(2015){Casey}, {Cooray}, {Capak}, {Fu}, {Kovac},
  {Lilly}, {Sanders}, {Scoville}, \& {Treister}}]{casey15}
{Casey}, C.~M., {Cooray}, A., {Capak}, P., {et~al.} 2015, \apjl, 808, L33,
  \dodoi{10.1088/2041-8205/808/2/L33}

\bibitem[{{Chiang} {et~al.}(2013){Chiang}, {Overzier}, \&
  {Gebhardt}}]{chiang13}
{Chiang}, Y.-K., {Overzier}, R., \& {Gebhardt}, K. 2013, \apj, 779, 127,
  \dodoi{10.1088/0004-637X/779/2/127}

\bibitem[{{Chiang} {et~al.}(2014){Chiang}, {Overzier}, \&
  {Gebhardt}}]{chiang14}
---. 2014, \apjl, 782, L3, \dodoi{10.1088/2041-8205/782/1/L3}

\bibitem[{{Chiang} {et~al.}(2017){Chiang}, {Overzier}, {Gebhardt}, \&
  {Henriques}}]{chiang17}
{Chiang}, Y.-K., {Overzier}, R.~A., {Gebhardt}, K., \& {Henriques}, B. 2017,
  \apjl, 844, L23, \dodoi{10.3847/2041-8213/aa7e7b}

\bibitem[{{Chiang} {et~al.}(2015){Chiang}, {Overzier}, {Gebhardt},
  {Finkelstein}, {Chiang}, {Hill}, {Blanc}, {Drory}, {Chonis}, {Zeimann},
  {Hagen}, {Schneider}, {Jogee}, {Ciardullo}, \& {Gronwall}}]{chiang15}
{Chiang}, Y.-K., {Overzier}, R.~A., {Gebhardt}, K., {et~al.} 2015, \apj, 808,
  37, \dodoi{10.1088/0004-637X/808/1/37}

\bibitem[{{Chiappetti} {et~al.}(2005){Chiappetti}, {Tajer}, {Trinchieri},
  {Maccagni}, {Maraschi}, {Paioro}, {Pierre}, {Surdej}, {Garcet}, {Gosset}, {Le
  F{\`e}vre}, {Bertin}, {McCracken}, {Mellier}, {Foucaud}, {Radovich},
  {Ripepi}, \& {Arnaboldi}}]{chiappetti05}
{Chiappetti}, L., {Tajer}, M., {Trinchieri}, G., {et~al.} 2005, \aap, 439, 413,
  \dodoi{10.1051/0004-6361:20042583}

\bibitem[{{Ciardullo} {et~al.}(2013){Ciardullo}, {Gronwall}, {Adams}, {Blanc},
  {Gebhardt}, {Finkelstein}, {Jogee}, {Hill}, {Drory}, {Hopp}, {Schneider},
  {Zeimann}, \& {Dalton}}]{ciardullo13}
{Ciardullo}, R., {Gronwall}, C., {Adams}, J.~J., {et~al.} 2013, \apj, 769, 83,
  \dodoi{10.1088/0004-637X/769/1/83}

\bibitem[{{Civano} {et~al.}(2016){Civano}, {Marchesi}, {Comastri}, {Urry},
  {Elvis}, {Cappelluti}, {Puccetti}, {Brusa}, {Zamorani}, {Hasinger},
  {Aldcroft}, {Alexand er}, {Allevato}, {Brunner}, {Capak}, {Finoguenov},
  {Fiore}, {Fruscione}, {Gilli}, {Glotfelty}, {Griffiths}, {Hao}, {Harrison},
  {Jahnke}, {Kartaltepe}, {Karim}, {LaMassa}, {Lanzuisi}, {Miyaji}, {Ranalli},
  {Salvato}, {Sargent}, {Scoville}, {Schawinski}, {Schinnerer}, {Silverman},
  {Smolcic}, {Stern}, {Toft}, {Trakhtenbrot}, {Treister}, \&
  {Vignali}}]{civano16}
{Civano}, F., {Marchesi}, S., {Comastri}, A., {et~al.} 2016, \apj, 819, 62,
  \dodoi{10.3847/0004-637X/819/1/62}

\bibitem[{{Cooke} {et~al.}(2014){Cooke}, {Hatch}, {Muldrew}, {Rigby}, \&
  {Kurk}}]{cooke14}
{Cooke}, E.~A., {Hatch}, N.~A., {Muldrew}, S.~I., {Rigby}, E.~E., \& {Kurk},
  J.~D. 2014, \mnras, 440, 3262, \dodoi{10.1093/mnras/stu522}

\bibitem[{{Cooper} {et~al.}(2008){Cooper}, {Newman}, {Weiner}, {Yan},
  {Willmer}, {Bundy}, {Coil}, {Conselice}, {Davis}, {Faber}, {Gerke},
  {Guhathakurta}, {Koo}, \& {Noeske}}]{cooper08}
{Cooper}, M.~C., {Newman}, J.~A., {Weiner}, B.~J., {et~al.} 2008, \mnras, 383,
  1058, \dodoi{10.1111/j.1365-2966.2007.12613.x}

\bibitem[{{Coppin} {et~al.}(2010){Coppin}, {Pope}, {Men{\'e}ndez-Delmestre},
  {Alexander}, {Dunlop}, {Egami}, {Gabor}, {Ibar}, {Ivison}, {Austermann},
  {Blain}, {Chapman}, {Clements}, {Dunne}, {Dye}, {Farrah}, {Hughes},
  {Mortier}, {Page}, {Rowan-Robinson}, {Scott}, {Simpson}, {Smail}, {Swinbank},
  {Vaccari}, \& {Yun}}]{coppin10}
{Coppin}, K., {Pope}, A., {Men{\'e}ndez-Delmestre}, K., {et~al.} 2010, \apj,
  713, 503, \dodoi{10.1088/0004-637X/713/1/503}

\bibitem[{{Cucciati} {et~al.}(2014){Cucciati}, {Zamorani}, {Lemaux},
  {Bardelli}, {Cimatti}, {Le F{\`e}vre}, {Cassata}, {Garilli}, {Le Brun},
  {Maccagni}, {Pentericci}, {Tasca}, {Thomas}, {Vanzella}, {Zucca}, {Amorin},
  {Capak}, {Cassar{\`a}}, {Castellano}, {Cuby}, {de la Torre}, {Durkalec},
  {Fontana}, {Giavalisco}, {Grazian}, {Hathi}, {Ilbert}, {Moreau}, {Paltani},
  {Ribeiro}, {Salvato}, {Schaerer}, {Scodeggio}, {Sommariva}, {Talia},
  {Taniguchi}, {Tresse}, {Vergani}, {Wang}, {Charlot}, {Contini}, {Fotopoulou},
  {L{\'o}pez-Sanjuan}, {Mellier}, \& {Scoville}}]{cucciati14}
{Cucciati}, O., {Zamorani}, G., {Lemaux}, B.~C., {et~al.} 2014, \aap, 570, A16,
  \dodoi{10.1051/0004-6361/201423811}

\bibitem[{{Cucciati} {et~al.}(2018){Cucciati}, {Lemaux}, {Zamorani}, {Le
  F{\`e}vre}, {Tasca}, {Hathi}, {Lee}, {Bardelli}, {Cassata}, {Garilli}, {Le
  Brun}, {Maccagni}, {Pentericci}, {Thomas}, {Vanzella}, {Zucca}, {Lubin},
  {Amorin}, {Cassar{\`a}}, {Cimatti}, {Talia}, {Vergani}, {Koekemoer}, {Pforr},
  \& {Salvato}}]{cucciati18}
{Cucciati}, O., {Lemaux}, B.~C., {Zamorani}, G., {et~al.} 2018, \aap, 619, A49,
  \dodoi{10.1051/0004-6361/201833655}

\bibitem[{{Daddi} {et~al.}(2021){Daddi}, {Valentino}, {Rich}, {Neill},
  {Gronke}, {O'Sullivan}, {Elbaz}, {Bournaud}, {Finoguenov}, {Marchal},
  {Delvecchio}, {Jin}, {Liu}, {Strazzullo}, {Calabro}, {Coogan}, {D'Eugenio},
  {Gobat}, {Kalita}, {Laursen}, {Martin}, {Puglisi}, {Schinnerer}, \&
  {Wang}}]{daddi21}
{Daddi}, E., {Valentino}, F., {Rich}, R.~M., {et~al.} 2021, \aap, 649, A78,
  \dodoi{10.1051/0004-6361/202038700}

\bibitem[{{Daddi} {et~al.}(2022){Daddi}, {Rich}, {Valentino}, {Jin},
  {Delvecchio}, {Liu}, {Strazzullo}, {Neill}, {Gobat}, {Finoguenov},
  {Bournaud}, {Elbaz}, {Kalita}, {O'Sullivan}, \& {Wang}}]{daddi22}
{Daddi}, E., {Rich}, R.~M., {Valentino}, F., {et~al.} 2022, arXiv e-prints,
  arXiv:2202.03715.
\newblock \doarXiv{2202.03715}

\bibitem[{{Dey} {et~al.}(2016){Dey}, {Lee}, {Reddy}, {Cooper}, {Inami}, {Hong},
  {Gonzalez}, \& {Jannuzi}}]{dey16}
{Dey}, A., {Lee}, K.-S., {Reddy}, N., {et~al.} 2016, \apj, 823, 11,
  \dodoi{10.3847/0004-637X/823/1/11}

\bibitem[{{Diener} {et~al.}(2015){Diener}, {Lilly}, {Ledoux}, {Zamorani},
  {Bolzonella}, {Murphy}, {Capak}, {Ilbert}, \& {McCracken}}]{diener15}
{Diener}, C., {Lilly}, S.~J., {Ledoux}, C., {et~al.} 2015, \apj, 802, 31,
  \dodoi{10.1088/0004-637X/802/1/31}

\bibitem[{{Digby-North} {et~al.}(2010){Digby-North}, {Nandra}, {Laird},
  {Steidel}, {Georgakakis}, {Bogosavljevi{\'c}}, {Erb}, {Shapley}, {Reddy}, \&
  {Aird}}]{digby10}
{Digby-North}, J.~A., {Nandra}, K., {Laird}, E.~S., {et~al.} 2010, \mnras, 407,
  846, \dodoi{10.1111/j.1365-2966.2010.16977.x}

\bibitem[{{Eisenhardt} {et~al.}(2008){Eisenhardt}, {Brodwin}, {Gonzalez},
  {Stanford}, {Stern}, {Barmby}, {Brown}, {Dawson}, {Dey}, {Doi}, {Galametz},
  {Jannuzi}, {Kochanek}, {Meyers}, {Morokuma}, \& {Moustakas}}]{eisenhardt08}
{Eisenhardt}, P. R.~M., {Brodwin}, M., {Gonzalez}, A.~H., {et~al.} 2008, \apj,
  684, 905, \dodoi{10.1086/590105}

\bibitem[{{Fakhouri} \& {Ma}(2009)}]{fakhouri09}
{Fakhouri}, O., \& {Ma}, C.-P. 2009, \mnras, 394, 1825,
  \dodoi{10.1111/j.1365-2966.2009.14480.x}

\bibitem[{{Faucher-Gigu{\`e}re} {et~al.}(2008){Faucher-Gigu{\`e}re},
  {Prochaska}, {Lidz}, {Hernquist}, \& {Zaldarriaga}}]{faucher08}
{Faucher-Gigu{\`e}re}, C.-A., {Prochaska}, J.~X., {Lidz}, A., {Hernquist}, L.,
  \& {Zaldarriaga}, M. 2008, \apj, 681, 831, \dodoi{10.1086/588648}

\bibitem[{{Flaugher} {et~al.}(2015){Flaugher}, {Diehl}, {Honscheid}, {Abbott},
  {Alvarez}, {Angstadt}, {Annis}, {Antonik}, {Ballester}, {Beaufore},
  {Bernstein}, {Bernstein}, {Bigelow}, {Bonati}, {Boprie}, {Brooks},
  {Buckley-Geer}, {Campa}, {Cardiel-Sas}, {Castander}, {Castilla}, {Cease},
  {Cela-Ruiz}, {Chappa}, {Chi}, {Cooper}, {da Costa}, {Dede}, {Derylo},
  {DePoy}, {de Vicente}, {Doel}, {Drlica-Wagner}, {Eiting}, {Elliott}, {Emes},
  {Estrada}, {Fausti Neto}, {Finley}, {Flores}, {Frieman}, {Gerdes},
  {Gladders}, {Gregory}, {Gutierrez}, {Hao}, {Holland}, {Holm}, {Huffman},
  {Jackson}, {James}, {Jonas}, {Karcher}, {Karliner}, {Kent}, {Kessler},
  {Kozlovsky}, {Kron}, {Kubik}, {Kuehn}, {Kuhlmann}, {Kuk}, {Lahav}, {Lathrop},
  {Lee}, {Levi}, {Lewis}, {Li}, {Mandrichenko}, {Marshall}, {Martinez},
  {Merritt}, {Miquel}, {Mu{\~n}oz}, {Neilsen}, {Nichol}, {Nord}, {Ogando},
  {Olsen}, {Palaio}, {Patton}, {Peoples}, {Plazas}, {Rauch}, {Reil}, {Rheault},
  {Roe}, {Rogers}, {Roodman}, {Sanchez}, {Scarpine}, {Schindler}, {Schmidt},
  {Schmitt}, {Schubnell}, {Schultz}, {Schurter}, {Scott}, {Serrano}, {Shaw},
  {Smith}, {Soares-Santos}, {Stefanik}, {Stuermer}, {Suchyta}, {Sypniewski},
  {Tarle}, {Thaler}, {Tighe}, {Tran}, {Tucker}, {Walker}, {Wang}, {Watson},
  {Weaverdyck}, {Wester}, {Woods}, {Yanny}, \& {DES Collaboration}}]{decam}
{Flaugher}, B., {Diehl}, H.~T., {Honscheid}, K., {et~al.} 2015, \aj, 150, 150,
  \dodoi{10.1088/0004-6256/150/5/150}

\bibitem[{{Gaia Collaboration} {et~al.}(2016){Gaia Collaboration}, {Prusti},
  {de Bruijne}, {Brown}, {Vallenari}, {Babusiaux}, {Bailer-Jones}, {Bastian},
  {Biermann}, {Evans}, \& et~al.}]{gaia1}
{Gaia Collaboration}, {Prusti}, T., {de Bruijne}, J.~H.~J., {et~al.} 2016,
  \aap, 595, A1, \dodoi{10.1051/0004-6361/201629272}

\bibitem[{{Gawiser} {et~al.}(2006){Gawiser}, {van Dokkum}, {Gronwall},
  {Ciardullo}, {Blanc}, {Castander}, {Feldmeier}, {Francke}, {Franx},
  {Haberzettl}, {Herrera}, {Hickey}, {Infante}, {Lira}, {Maza}, {Quadri},
  {Richardson}, {Schawinski}, {Schirmer}, {Taylor}, {Treister}, {Urry}, \&
  {Virani}}]{gawiser06}
{Gawiser}, E., {van Dokkum}, P.~G., {Gronwall}, C., {et~al.} 2006, \apjl, 642,
  L13, \dodoi{10.1086/504467}

\bibitem[{{Gawiser} {et~al.}(2007){Gawiser}, {Francke}, {Lai}, {Schawinski},
  {Gronwall}, {Ciardullo}, {Quadri}, {Orsi}, {Barrientos}, {Blanc}, {Fazio},
  {Feldmeier}, {Huang}, {Infante}, {Lira}, {Padilla}, {Taylor}, {Treister},
  {Urry}, {van Dokkum}, \& {Virani}}]{gawiser07}
{Gawiser}, E., {Francke}, H., {Lai}, K., {et~al.} 2007, \apj, 671, 278,
  \dodoi{10.1086/522955}

\bibitem[{{Gisler}(1978)}]{gisler78}
{Gisler}, G.~R. 1978, \mnras, 183, 633, \dodoi{10.1093/mnras/183.4.633}

\bibitem[{{Gopu} {et~al.}(2014){Gopu}, {Hayashi}, {Young}, {Harbeck},
  {Boroson}, {Liu}, {Kotulla}, {Shaw}, {Henschel}, {Rajagopal}, {Stobie},
  {Knezek}, {Martin}, \& {Archbold}}]{gopu14}
{Gopu}, A., {Hayashi}, S., {Young}, M.~D., {et~al.} 2014, in \procspie, Vol.
  9152, Software and Cyberinfrastructure for Astronomy III, 91520E,
  \dodoi{10.1117/12.2057123}

\bibitem[{{Gouin} {et~al.}(2022){Gouin}, {Aghanim}, {Dole}, {Polletta}, \&
  {Park}}]{gouin22}
{Gouin}, C., {Aghanim}, N., {Dole}, H., {Polletta}, M., \& {Park}, C. 2022,
  arXiv e-prints, arXiv:2203.16276.
\newblock \doarXiv{2203.16276}

\bibitem[{{Gronwall} {et~al.}(2007){Gronwall}, {Ciardullo}, {Hickey},
  {Gawiser}, {Feldmeier}, {van Dokkum}, {Urry}, {Herrera}, {Lehmer}, {Infante},
  {Orsi}, {Marchesini}, {Blanc}, {Francke}, {Lira}, \& {Treister}}]{gronwall07}
{Gronwall}, C., {Ciardullo}, R., {Hickey}, T., {et~al.} 2007, \apj, 667, 79,
  \dodoi{10.1086/520324}

\bibitem[{{Guaita} {et~al.}(2010){Guaita}, {Gawiser}, {Padilla}, {Francke},
  {Bond}, {Gronwall}, {Ciardullo}, {Feldmeier}, {Sinawa}, {Blanc}, \&
  {Virani}}]{guaita10}
{Guaita}, L., {Gawiser}, E., {Padilla}, N., {et~al.} 2010, \apj, 714, 255,
  \dodoi{10.1088/0004-637X/714/1/255}

\bibitem[{{Guaita} {et~al.}(2011){Guaita}, {Acquaviva}, {Padilla}, {Gawiser},
  {Bond}, {Ciardullo}, {Treister}, {Kurczynski}, {Gronwall}, {Lira}, \&
  {Schawinski}}]{guaita11}
{Guaita}, L., {Acquaviva}, V., {Padilla}, N., {et~al.} 2011, \apj, 733, 114,
  \dodoi{10.1088/0004-637X/733/2/114}

\bibitem[{{Harbeck} {et~al.}(2014){Harbeck}, {Boroson}, {Lesser}, {Rajagopal},
  {Yeatts}, {Corson}, {Liu}, {Dell'Antonio}, {Kotulla}, {Ouellette}, {Hooper},
  {Smith}, {Bredthauer}, {Martin}, {Muller}, {Knezek}, \& {Hunten}}]{odi_old}
{Harbeck}, D.~R., {Boroson}, T., {Lesser}, M., {et~al.} 2014, in Society of
  Photo-Optical Instrumentation Engineers (SPIE) Conference Series, Vol. 9147,
  Ground-based and Airborne Instrumentation for Astronomy V, ed. S.~K.
  {Ramsay}, I.~S. {McLean}, \& H.~{Takami}, 91470P, \dodoi{10.1117/12.2056651}

\bibitem[{{Harbeck} {et~al.}(2018){Harbeck}, {Lesser}, {Liu}, {Stupak},
  {George}, {Harris}, {Poczulp}, {Rajagopal}, {Kotulla}, {Ouellete}, {Hooper},
  {Smith}, {Mason}, {Onaka}, {Chin}, {Hunting}, \& {Christensen}}]{odi_new}
{Harbeck}, D.~R., {Lesser}, M., {Liu}, W., {et~al.} 2018, in Society of
  Photo-Optical Instrumentation Engineers (SPIE) Conference Series, Vol. 10702,
  Ground-based and Airborne Instrumentation for Astronomy VII, ed. C.~J.
  {Evans}, L.~{Simard}, \& H.~{Takami}, 1070229, \dodoi{10.1117/12.2311528}

\bibitem[{{Harikane} {et~al.}(2019){Harikane}, {Ouchi}, {Ono}, {Fujimoto},
  {Donevski}, {Shibuya}, {Faisst}, {Goto}, {Hatsukade}, {Kashikawa}, {Kohno},
  {Hashimoto}, {Higuchi}, {Inoue}, {Lin}, {Martin}, {Overzier}, {Smail},
  {Toshikawa}, {Umehata}, {Ao}, {Chapman}, {Clements}, {Im}, {Jing},
  {Kawaguchi}, {Lee}, {Lee}, {Lin}, {Matsuoka}, {Marinello}, {Nagao},
  {Onodera}, {Toft}, \& {Wang}}]{harikane19}
{Harikane}, Y., {Ouchi}, M., {Ono}, Y., {et~al.} 2019, \apj, 883, 142,
  \dodoi{10.3847/1538-4357/ab2cd5}

\bibitem[{{Hatch} {et~al.}(2011){Hatch}, {Kurk}, {Pentericci}, {Venemans},
  {Kuiper}, {Miley}, \& {R{\"o}ttgering}}]{hatch11}
{Hatch}, N.~A., {Kurk}, J.~D., {Pentericci}, L., {et~al.} 2011, \mnras, 415,
  2993, \dodoi{10.1111/j.1365-2966.2011.18735.x}

\bibitem[{{Hayashi} {et~al.}(2012){Hayashi}, {Kodama}, {Tadaki}, {Koyama}, \&
  {Tanaka}}]{hayashi12}
{Hayashi}, M., {Kodama}, T., {Tadaki}, K.-i., {Koyama}, Y., \& {Tanaka}, I.
  2012, \apj, 757, 15, \dodoi{10.1088/0004-637X/757/1/15}

\bibitem[{{Hayashino} {et~al.}(2004){Hayashino}, {Matsuda}, {Tamura},
  {Yamauchi}, {Yamada}, {Ajiki}, {Fujita}, {Murayama}, {Nagao}, {Ohta},
  {Okamura}, {Ouchi}, {Shimasaku}, {Shioya}, \& {Taniguchi}}]{hayashino04}
{Hayashino}, T., {Matsuda}, Y., {Tamura}, H., {et~al.} 2004, \aj, 128, 2073,
  \dodoi{10.1086/424935}

\bibitem[{{Hennawi} {et~al.}(2015){Hennawi}, {Prochaska}, {Cantalupo}, \&
  {Arrigoni-Battaia}}]{hennawi15}
{Hennawi}, J.~F., {Prochaska}, J.~X., {Cantalupo}, S., \& {Arrigoni-Battaia},
  F. 2015, Science, 348, 779, \dodoi{10.1126/science.aaa5397}

\bibitem[{{Hine} {et~al.}(2016){Hine}, {Geach}, {Alexander}, {Lehmer},
  {Chapman}, \& {Matsuda}}]{hine15}
{Hine}, N.~K., {Geach}, J.~E., {Alexander}, D.~M., {et~al.} 2016, \mnras, 455,
  2363, \dodoi{10.1093/mnras/stv2448}

\bibitem[{{Hopkins} {et~al.}(2014){Hopkins}, {Kere{\v{s}}}, {O{\~n}orbe},
  {Faucher-Gigu{\`e}re}, {Quataert}, {Murray}, \& {Bullock}}]{hopkins14}
{Hopkins}, P.~F., {Kere{\v{s}}}, D., {O{\~n}orbe}, J., {et~al.} 2014, \mnras,
  445, 581, \dodoi{10.1093/mnras/stu1738}

\bibitem[{{Hu} {et~al.}(2021){Hu}, {Wang}, {Infante}, {Rhoads}, {Zheng},
  {Yang}, {Malhotra}, {Barrientos}, {Jiang}, {Gonz{\'a}lez-L{\'o}pez},
  {Prieto}, {Perez}, {Hibon}, {Galaz}, {Coughlin}, {Harish}, {Kong}, {Kang},
  {Khostovan}, {Pharo}, {Valdes}, {Wold}, {Walker}, \& {Zheng}}]{hu21}
{Hu}, W., {Wang}, J., {Infante}, L., {et~al.} 2021, Nature Astronomy, 5, 485,
  \dodoi{10.1038/s41550-020-01291-y}

\bibitem[{{Intema} {et~al.}(2006){Intema}, {Venemans}, {Kurk}, {Ouchi},
  {Kodama}, {R{\"o}ttgering}, {Miley}, \& {Overzier}}]{intema06}
{Intema}, H.~T., {Venemans}, B.~P., {Kurk}, J.~D., {et~al.} 2006, \aap, 456,
  433, \dodoi{10.1051/0004-6361:20064812}

\bibitem[{{Kashikawa} {et~al.}(2007){Kashikawa}, {Kitayama}, {Doi}, {Misawa},
  {Komiyama}, \& {Ota}}]{kashikawa07}
{Kashikawa}, N., {Kitayama}, T., {Doi}, M., {et~al.} 2007, \apj, 663, 765,
  \dodoi{10.1086/518410}

\bibitem[{{Komiyama} {et~al.}(2018){Komiyama}, {Obuchi}, {Nakaya}, {Kamata},
  {Kawanomoto}, {Utsumi}, {Miyazaki}, {Uraguchi}, {Furusawa}, {Morokuma},
  {Uchida}, {Miyatake}, {Mineo}, {Fujimori}, {Aihara}, {Karoji}, {Gunn}, \&
  {Wang}}]{komiyama18}
{Komiyama}, Y., {Obuchi}, Y., {Nakaya}, H., {et~al.} 2018, \pasj, 70, S2,
  \dodoi{10.1093/pasj/psx069}

\bibitem[{{Koyama} {et~al.}(2013){Koyama}, {Smail}, {Kurk}, {Geach}, {Sobral},
  {Kodama}, {Nakata}, {Swinbank}, {Best}, {Hayashi}, \& {Tadaki}}]{koyama13b}
{Koyama}, Y., {Smail}, I., {Kurk}, J., {et~al.} 2013, \mnras, 434, 423,
  \dodoi{10.1093/mnras/stt1035}

\bibitem[{{Krishnan} {et~al.}(2017){Krishnan}, {Hatch}, {Almaini}, {Kocevski},
  {Cooke}, {Hartley}, {Hasinger}, {Maltby}, {Muldrew}, \&
  {Simpson}}]{krishnan17}
{Krishnan}, C., {Hatch}, N.~A., {Almaini}, O., {et~al.} 2017, \mnras, 470,
  2170, \dodoi{10.1093/mnras/stx1315}

\bibitem[{{Kurk} {et~al.}(2004){Kurk}, {Pentericci}, {R{\"o}ttgering}, \&
  {Miley}}]{kurk04}
{Kurk}, J.~D., {Pentericci}, L., {R{\"o}ttgering}, H.~J.~A., \& {Miley}, G.~K.
  2004, \aap, 428, 793, \dodoi{10.1051/0004-6361:20040075}

\bibitem[{{Kusakabe} {et~al.}(2018){Kusakabe}, {Shimasaku}, {Ouchi},
  {Nakajima}, {Goto}, {Hashimoto}, {Konno}, {Harikane}, {Silverman}, \&
  {Capak}}]{kusakabe18}
{Kusakabe}, H., {Shimasaku}, K., {Ouchi}, M., {et~al.} 2018, \pasj, 70, 4,
  \dodoi{10.1093/pasj/psx148}

\bibitem[{{Laigle} {et~al.}(2016){Laigle}, {McCracken}, {Ilbert}, {Hsieh},
  {Davidzon}, {Capak}, {Hasinger}, {Silverman}, {Pichon}, {Coupon}, {Aussel},
  {Le Borgne}, {Caputi}, {Cassata}, {Chang}, {Civano}, {Dunlop}, {Fynbo},
  {Kartaltepe}, {Koekemoer}, {Le F{\`e}vre}, {Le Floc'h}, {Leauthaud}, {Lilly},
  {Lin}, {Marchesi}, {Milvang-Jensen}, {Salvato}, {Sanders}, {Scoville},
  {Smolcic}, {Stockmann}, {Taniguchi}, {Tasca}, {Toft}, {Vaccari}, \&
  {Zabl}}]{laigle16}
{Laigle}, C., {McCracken}, H.~J., {Ilbert}, O., {et~al.} 2016, \apjs, 224, 24,
  \dodoi{10.3847/0067-0049/224/2/24}

\bibitem[{{Le F{\`e}vre} {et~al.}(2015){Le F{\`e}vre}, {Tasca}, {Cassata},
  {Garilli}, {Le Brun}, {Maccagni}, {Pentericci}, {Thomas}, {Vanzella},
  {Zamorani}, {Zucca}, {Amorin}, {Bardelli}, {Capak}, {Cassar{\`a}},
  {Castellano}, {Cimatti}, {Cuby}, {Cucciati}, {de la Torre}, {Durkalec},
  {Fontana}, {Giavalisco}, {Grazian}, {Hathi}, {Ilbert}, {Lemaux}, {Moreau},
  {Paltani}, {Ribeiro}, {Salvato}, {Schaerer}, {Scodeggio}, {Sommariva},
  {Talia}, {Taniguchi}, {Tresse}, {Vergani}, {Wang}, {Charlot}, {Contini},
  {Fotopoulou}, {L{\'o}pez-Sanjuan}, {Mellier}, \& {Scoville}}]{lefevre15}
{Le F{\`e}vre}, O., {Tasca}, L.~A.~M., {Cassata}, P., {et~al.} 2015, \aap, 576,
  A79, \dodoi{10.1051/0004-6361/201423829}

\bibitem[{{Lee} {et~al.}(2014{\natexlab{a}}){Lee}, {Hennawi}, {Stark},
  {Prochaska}, {White}, {Schlegel}, {Eilers}, {Arinyo-i-Prats}, {Suzuki},
  {Croft}, {Caputi}, {Cassata}, {Ilbert}, {Garilli}, {Koekemoer}, {Le Brun},
  {Le F{\`e}vre}, {Maccagni}, {Nugent}, {Taniguchi}, {Tasca}, {Tresse},
  {Zamorani}, \& {Zucca}}]{lee14}
{Lee}, K.-G., {Hennawi}, J.~F., {Stark}, C., {et~al.} 2014{\natexlab{a}},
  \apjl, 795, L12, \dodoi{10.1088/2041-8205/795/1/L12}

\bibitem[{{Lee} {et~al.}(2016){Lee}, {Hennawi}, {White}, {Prochaska},
  {Font-Ribera}, {Schlegel}, {Rich}, {Suzuki}, {Stark}, {Le F{\`e}vre},
  {Nugent}, {Salvato}, \& {Zamorani}}]{lee16}
{Lee}, K.-G., {Hennawi}, J.~F., {White}, M., {et~al.} 2016, \apj, 817, 160,
  \dodoi{10.3847/0004-637X/817/2/160}

\bibitem[{{Lee} {et~al.}(2014{\natexlab{b}}){Lee}, {Dey}, {Hong}, {Reddy},
  {Wilson}, {Jannuzi}, {Inami}, \& {Gonzalez}}]{kslee14}
{Lee}, K.-S., {Dey}, A., {Hong}, S., {et~al.} 2014{\natexlab{b}}, \apj, 796,
  126, \dodoi{10.1088/0004-637X/796/2/126}

\bibitem[{{Lehmer} {et~al.}(2009){Lehmer}, {Alexander}, {Geach}, {Smail},
  {Basu-Zych}, {Bauer}, {Chapman}, {Matsuda}, {Scharf}, {Volonteri}, \&
  {Yamada}}]{lehmer09}
{Lehmer}, B.~D., {Alexander}, D.~M., {Geach}, J.~E., {et~al.} 2009, \apj, 691,
  687, \dodoi{10.1088/0004-637X/691/1/687}

\bibitem[{{Lemaux} {et~al.}(2014){Lemaux}, {Cucciati}, {Tasca}, {Le F{\`e}vre},
  {Zamorani}, {Cassata}, {Garilli}, {Le Brun}, {Maccagni}, {Pentericci},
  {Thomas}, {Vanzella}, {Zucca}, {Amor{\'\i}n}, {Bardelli}, {Capak},
  {Cassar{\`a}}, {Castellano}, {Cimatti}, {Cuby}, {de la Torre}, {Durkalec},
  {Fontana}, {Giavalisco}, {Grazian}, {Hathi}, {Ilbert}, {Moreau}, {Paltani},
  {Ribeiro}, {Salvato}, {Schaerer}, {Scodeggio}, {Sommariva}, {Talia},
  {Taniguchi}, {Tresse}, {Vergani}, {Wang}, {Charlot}, {Contini}, {Fotopoulou},
  {Gal}, {Kocevski}, {L{\'o}pez-Sanjuan}, {Lubin}, {Mellier}, {Sadibekova}, \&
  {Scoville}}]{lemaux14}
{Lemaux}, B.~C., {Cucciati}, O., {Tasca}, L.~A.~M., {et~al.} 2014, \aap, 572,
  A41, \dodoi{10.1051/0004-6361/201423828}

\bibitem[{{Lemaux} {et~al.}(2020){Lemaux}, {Cucciati}, {Le F{\`e}vre},
  {Zamorani}, {Lubin}, {Hathi}, {Ilbert}, {Pelliccia}, {Amor{\'\i}n},
  {Bardelli}, {Cassata}, {Gal}, {Garilli}, {Guaita}, {Giavalisco}, {Hung},
  {Koekemoer}, {Maccagni}, {Pentericci}, {Ribeiro}, {Schaerer}, {Shen},
  {Talia}, {Tomczak}, {Vanzella}, {Vergani}, \& {Zucca}}]{lemaux20}
{Lemaux}, B.~C., {Cucciati}, O., {Le F{\`e}vre}, O., {et~al.} 2020, arXiv
  e-prints, arXiv:2009.03324.
\newblock \doarXiv{2009.03324}

\bibitem[{{Lilly} {et~al.}(2007){Lilly}, {Le F{\`e}vre}, {Renzini}, {Zamorani},
  {Scodeggio}, {Contini}, {Carollo}, {Hasinger}, {Kneib}, {Iovino}, {Le Brun},
  {Maier}, {Mainieri}, {Mignoli}, {Silverman}, {Tasca}, {Bolzonella},
  {Bongiorno}, {Bottini}, {Capak}, {Caputi}, {Cimatti}, {Cucciati}, {Daddi},
  {Feldmann}, {Franzetti}, {Garilli}, {Guzzo}, {Ilbert}, {Kampczyk}, {Kovac},
  {Lamareille}, {Leauthaud}, {Le Borgne}, {McCracken}, {Marinoni}, {Pello},
  {Ricciardelli}, {Scarlata}, {Vergani}, {Sanders}, {Schinnerer}, {Scoville},
  {Taniguchi}, {Arnouts}, {Aussel}, {Bardelli}, {Brusa}, {Cappi}, {Ciliegi},
  {Finoguenov}, {Foucaud}, {Franceschini}, {Halliday}, {Impey}, {Knobel},
  {Koekemoer}, {Kurk}, {Maccagni}, {Maddox}, {Marano}, {Marconi}, {Meneux},
  {Mobasher}, {Moreau}, {Peacock}, {Porciani}, {Pozzetti}, {Scaramella},
  {Schiminovich}, {Shopbell}, {Smail}, {Thompson}, {Tresse}, {Vettolani},
  {Zanichelli}, \& {Zucca}}]{lilly07}
{Lilly}, S.~J., {Le F{\`e}vre}, O., {Renzini}, A., {et~al.} 2007, \apjs, 172,
  70, \dodoi{10.1086/516589}

\bibitem[{{Lilly} {et~al.}(2009){Lilly}, {Le Brun}, {Maier}, {Mainieri},
  {Mignoli}, {Scodeggio}, {Zamorani}, {Carollo}, {Contini}, {Kneib}, {Le
  F{\`e}vre}, {Renzini}, {Bardelli}, {Bolzonella}, {Bongiorno}, {Caputi},
  {Coppa}, {Cucciati}, {de la Torre}, {de Ravel}, {Franzetti}, {Garilli},
  {Iovino}, {Kampczyk}, {Kovac}, {Knobel}, {Lamareille}, {Le Borgne}, {Pello},
  {Peng}, {P{\'e}rez-Montero}, {Ricciardelli}, {Silverman}, {Tanaka}, {Tasca},
  {Tresse}, {Vergani}, {Zucca}, {Ilbert}, {Salvato}, {Oesch}, {Abbas},
  {Bottini}, {Capak}, {Cappi}, {Cassata}, {Cimatti}, {Elvis}, {Fumana},
  {Guzzo}, {Hasinger}, {Koekemoer}, {Leauthaud}, {Maccagni}, {Marinoni},
  {McCracken}, {Memeo}, {Meneux}, {Porciani}, {Pozzetti}, {Sanders},
  {Scaramella}, {Scarlata}, {Scoville}, {Shopbell}, \& {Taniguchi}}]{lilly09}
{Lilly}, S.~J., {Le Brun}, V., {Maier}, C., {et~al.} 2009, \apjs, 184, 218,
  \dodoi{10.1088/0067-0049/184/2/218}

\bibitem[{{Lim} {et~al.}(2021){Lim}, {Scott}, {Babul}, {Barnes}, {Kay},
  {McCarthy}, {Rennehan}, \& {Vogelsberger}}]{lim21}
{Lim}, S., {Scott}, D., {Babul}, A., {et~al.} 2021, \mnras, 501, 1803,
  \dodoi{10.1093/mnras/staa3693}

\bibitem[{{Lotz} {et~al.}(2013){Lotz}, {Papovich}, {Faber}, {Ferguson},
  {Grogin}, {Guo}, {Kocevski}, {Koekemoer}, {Lee}, {McIntosh}, {Momcheva},
  {Rudnick}, {Saintonge}, {Tran}, {van der Wel}, \& {Willmer}}]{lotz13}
{Lotz}, J.~M., {Papovich}, C., {Faber}, S.~M., {et~al.} 2013, \apj, 773, 154,
  \dodoi{10.1088/0004-637X/773/2/154}

\bibitem[{{Madau} \& {Dickinson}(2014)}]{madau14}
{Madau}, P., \& {Dickinson}, M. 2014, \araa, 52, 415,
  \dodoi{10.1146/annurev-astro-081811-125615}

\bibitem[{{Marchesi} {et~al.}(2016){Marchesi}, {Civano}, {Elvis}, {Salvato},
  {Brusa}, {Comastri}, {Gilli}, {Hasinger}, {Lanzuisi}, {Miyaji}, {Treister},
  {Urry}, {Vignali}, {Zamorani}, {Allevato}, {Cappelluti}, {Cardamone},
  {Finoguenov}, {Griffiths}, {Karim}, {Laigle}, {LaMassa}, {Jahnke}, {Ranalli},
  {Schawinski}, {Schinnerer}, {Silverman}, {Smolcic}, {Suh}, \&
  {Trakhtenbrot}}]{marchesi16}
{Marchesi}, S., {Civano}, F., {Elvis}, M., {et~al.} 2016, \apj, 817, 34,
  \dodoi{10.3847/0004-637X/817/1/34}

\bibitem[{{Mart{\'\i}n-Navarro} {et~al.}(2018){Mart{\'\i}n-Navarro},
  {Vazdekis}, {Falc{\'o}n-Barroso}, {La Barbera}, {Y{\i}ld{\i}r{\i}m}, \& {van
  de Ven}}]{martin18}
{Mart{\'\i}n-Navarro}, I., {Vazdekis}, A., {Falc{\'o}n-Barroso}, J., {et~al.}
  2018, \mnras, 475, 3700, \dodoi{10.1093/mnras/stx3346}

\bibitem[{{Matsuda} {et~al.}(2004){Matsuda}, {Yamada}, {Hayashino}, {Tamura},
  {Yamauchi}, {Ajiki}, {Fujita}, {Murayama}, {Nagao}, {Ohta}, {Okamura},
  {Ouchi}, {Shimasaku}, {Shioya}, \& {Taniguchi}}]{matsuda04}
{Matsuda}, Y., {Yamada}, T., {Hayashino}, T., {et~al.} 2004, \aj, 128, 569,
  \dodoi{10.1086/422020}

\bibitem[{{Matsuda} {et~al.}(2005){Matsuda}, {Yamada}, {Hayashino}, {Tamura},
  {Yamauchi}, {Murayama}, {Nagao}, {Ohta}, {Okamura}, {Ouchi}, {Shimasaku},
  {Shioya}, \& {Taniguchi}}]{matsuda05}
---. 2005, \apjl, 634, L125, \dodoi{10.1086/499071}

\bibitem[{{McConachie} {et~al.}(2022){McConachie}, {Wilson}, {Forrest},
  {Marsan}, {Muzzin}, {Cooper}, {Annunziatella}, {Marchesini}, {Chan}, {Gomez},
  {Abdullah}, {Saracco}, \& {Nantais}}]{mcconachie22}
{McConachie}, I., {Wilson}, G., {Forrest}, B., {et~al.} 2022, \apj, 926, 37,
  \dodoi{10.3847/1538-4357/ac2b9f}

\bibitem[{{Miyazaki} {et~al.}(2018){Miyazaki}, {Komiyama}, {Kawanomoto}, {Doi},
  {Furusawa}, {Hamana}, {Hayashi}, {Ikeda}, {Kamata}, {Karoji}, {Koike},
  {Kurakami}, {Miyama}, {Morokuma}, {Nakata}, {Namikawa}, {Nakaya}, {Nariai},
  {Obuchi}, {Oishi}, {Okada}, {Okura}, {Tait}, {Takata}, {Tanaka}, {Tanaka},
  {Terai}, {Tomono}, {Uraguchi}, {Usuda}, {Utsumi}, {Yamada}, {Yamanoi},
  {Aihara}, {Fujimori}, {Mineo}, {Miyatake}, {Oguri}, {Uchida}, {Tanaka},
  {Yasuda}, {Takada}, {Murayama}, {Nishizawa}, {Sugiyama}, {Chiba}, {Futamase},
  {Wang}, {Chen}, {Ho}, {Liaw}, {Chiu}, {Ho}, {Lai}, {Lee}, {Jeng}, {Iwamura},
  {Armstrong}, {Bickerton}, {Bosch}, {Gunn}, {Lupton}, {Loomis}, {Price},
  {Smith}, {Strauss}, {Turner}, {Suzuki}, {Miyazaki}, {Muramatsu}, {Yamamoto},
  {Endo}, {Ezaki}, {Ito}, {Kawaguchi}, {Sofuku}, {Taniike}, {Akutsu}, {Dojo},
  {Kasumi}, {Matsuda}, {Imoto}, {Miwa}, {Suzuki}, {Takeshi}, \&
  {Yokota}}]{miyazaki18}
{Miyazaki}, S., {Komiyama}, Y., {Kawanomoto}, S., {et~al.} 2018, \pasj, 70, S1,
  \dodoi{10.1093/pasj/psx063}

\bibitem[{{Mo} {et~al.}(2018){Mo}, {Gonzalez}, {Stern}, {Brodwin}, {Decker},
  {Eisenhardt}, {Moravec}, {Stanford}, \& {Wylezalek}}]{mo18}
{Mo}, W., {Gonzalez}, A., {Stern}, D., {et~al.} 2018, \apj, 869, 131,
  \dodoi{10.3847/1538-4357/aaef83}

\bibitem[{{Momose} {et~al.}(2019){Momose}, {Goto}, {Utsumi}, {Hashimoto},
  {Chiang}, {Kim}, {Kashikawa}, {Shimasaku}, \& {Miyazaki}}]{momose19}
{Momose}, R., {Goto}, T., {Utsumi}, Y., {et~al.} 2019, \mnras, 488, 120,
  \dodoi{10.1093/mnras/stz1707}

\bibitem[{{Momose} {et~al.}(2021){Momose}, {Shimasaku}, {Kashikawa},
  {Nagamine}, {Shimizu}, {Nakajima}, {Terao}, {Kusakabe}, {Ando}, {Motohara},
  \& {Spitler}}]{momose20}
{Momose}, R., {Shimasaku}, K., {Kashikawa}, N., {et~al.} 2021, \apj, 909, 117,
  \dodoi{10.3847/1538-4357/abd2af}

\bibitem[{{Monson} {et~al.}(2021){Monson}, {Lehmer}, {Doore}, {Eufrasio},
  {Bonine}, {Alexander}, {Harrison}, {Kubo}, {Mantha}, {Saez}, {Straughn}, \&
  {Umehata}}]{monson21}
{Monson}, E.~B., {Lehmer}, B.~D., {Doore}, K., {et~al.} 2021, \apj, 919, 51,
  \dodoi{10.3847/1538-4357/ac0f84}

\bibitem[{{Moutard} {et~al.}(2020){Moutard}, {Malavasi}, {Sawicki}, {Arnouts},
  \& {Tripathi}}]{moutard20}
{Moutard}, T., {Malavasi}, N., {Sawicki}, M., {Arnouts}, S., \& {Tripathi}, S.
  2020, \mnras, 495, 4237, \dodoi{10.1093/mnras/staa1434}

\bibitem[{{Muldrew} {et~al.}(2015){Muldrew}, {Hatch}, \& {Cooke}}]{muldrew15}
{Muldrew}, S.~I., {Hatch}, N.~A., \& {Cooke}, E.~A. 2015, \mnras, 452, 2528,
  \dodoi{10.1093/mnras/stv1449}

\bibitem[{{Nakajima} {et~al.}(2012){Nakajima}, {Ouchi}, {Shimasaku}, {Ono},
  {Lee}, {Foucaud}, {Ly}, {Dale}, {Salim}, {Finn}, {Almaini}, \&
  {Okamura}}]{nakajima12}
{Nakajima}, K., {Ouchi}, M., {Shimasaku}, K., {et~al.} 2012, \apj, 745, 12,
  \dodoi{10.1088/0004-637X/745/1/12}

\bibitem[{{Negrello} {et~al.}(2017){Negrello}, {Gonzalez-Nuevo}, {De Zotti},
  {Bonato}, {Cai}, {Clements}, {Danese}, {Dole}, {Greenslade}, {Lapi}, \&
  {Montier}}]{negrello17}
{Negrello}, M., {Gonzalez-Nuevo}, J., {De Zotti}, G., {et~al.} 2017, \mnras,
  470, 2253, \dodoi{10.1093/mnras/stx1367}

\bibitem[{{Newman} {et~al.}(2020){Newman}, {Rudie}, {Blanc}, {Kelson},
  {Rhoades}, {Hare}, {P{\'e}rez}, {Benson}, {Dressler}, {Gonzalez},
  {Kollmeier}, {Konidaris}, {Mulchaey}, {Rauch}, {Le F{\`e}vre}, {Lemaux},
  {Cucciati}, \& {Lilly}}]{newman20}
{Newman}, A.~B., {Rudie}, G.~C., {Blanc}, G.~A., {et~al.} 2020, \apj, 891, 147,
  \dodoi{10.3847/1538-4357/ab75ee}

\bibitem[{{Oke} \& {Gunn}(1983)}]{oke83}
{Oke}, J.~B., \& {Gunn}, J.~E. 1983, \apj, 266, 713, \dodoi{10.1086/160817}

\bibitem[{{Ouchi} {et~al.}(2020){Ouchi}, {Ono}, \& {Shibuya}}]{ouchi20}
{Ouchi}, M., {Ono}, Y., \& {Shibuya}, T. 2020, \araa, 58, 617,
  \dodoi{10.1146/annurev-astro-032620-021859}

\bibitem[{{Ouchi} {et~al.}(2008){Ouchi}, {Shimasaku}, {Akiyama}, {Simpson},
  {Saito}, {Ueda}, {Furusawa}, {Sekiguchi}, {Yamada}, {Kodama}, {Kashikawa},
  {Okamura}, {Iye}, {Takata}, {Yoshida}, \& {Yoshida}}]{ouchi08}
{Ouchi}, M., {Shimasaku}, K., {Akiyama}, M., {et~al.} 2008, \apjs, 176, 301,
  \dodoi{10.1086/527673}

\bibitem[{{Overzier}(2016)}]{overzier16}
{Overzier}, R.~A. 2016, \aapr, 24, 14, \dodoi{10.1007/s00159-016-0100-3}

\bibitem[{{Overzier} {et~al.}(2006){Overzier}, {Bouwens}, {Illingworth}, \&
  {Franx}}]{overzier06}
{Overzier}, R.~A., {Bouwens}, R.~J., {Illingworth}, G.~D., \& {Franx}, M. 2006,
  \apjl, 648, L5, \dodoi{10.1086/507678}

\bibitem[{{Overzier} {et~al.}(2013){Overzier}, {Nesvadba}, {Dijkstra}, {Hatch},
  {Lehnert}, {Villar-Mart{\'\i}n}, {Wilman}, \& {Zirm}}]{overzier13}
{Overzier}, R.~A., {Nesvadba}, N.~P.~H., {Dijkstra}, M., {et~al.} 2013, \apj,
  771, 89, \dodoi{10.1088/0004-637X/771/2/89}

\bibitem[{{Penny} {et~al.}(2018){Penny}, {Masters}, {Smethurst}, {Nichol},
  {Krawczyk}, {Bizyaev}, {Greene}, {Liu}, {Marinelli}, {Rembold}, {Riffel},
  {Ilha}, {Wylezalek}, {Andrews}, {Bundy}, {Drory}, {Oravetz}, \&
  {Pan}}]{penny18}
{Penny}, S.~J., {Masters}, K.~L., {Smethurst}, R., {et~al.} 2018, \mnras, 476,
  979, \dodoi{10.1093/mnras/sty202}

\bibitem[{{Planck Collaboration} {et~al.}(2014){Planck Collaboration}, {Ade},
  {Aghanim}, {Arg{\"u}eso}, {Armitage-Caplan}, {Arnaud}, {Ashdown},
  {Atrio-Barandela}, {Aumont}, {Baccigalupi}, {Banday}, {Barreiro}, {Bartlett},
  {Battaner}, {Beelen}, {Benabed}, {Beno{\^\i}t}, {Benoit-L{\'e}vy}, {Bernard},
  {Bersanelli}, {Bielewicz}, {Bobin}, {Bock}, {Bonaldi}, {Bonavera}, {Bond},
  {Borrill}, {Bouchet}, {Bridges}, {Bucher}, {Burigana}, {Butler}, {Cardoso},
  {Carvalho}, {Catalano}, {Challinor}, {Chamballu}, {Chen}, {Chiang}, {Chiang},
  {Christensen}, {Church}, {Clemens}, {Clements}, {Colombi}, {Colombo},
  {Couchot}, {Coulais}, {Crill}, {Curto}, {Cuttaia}, {Danese}, {Davies},
  {Davis}, {de Bernardis}, {de Rosa}, {de Zotti}, {Delabrouille}, {Delouis},
  {D{\'e}sert}, {Dickinson}, {Diego}, {Dole}, {Donzelli}, {Dor{\'e}},
  {Douspis}, {Dupac}, {Efstathiou}, {En{\ss}lin}, {Eriksen}, {Finelli},
  {Forni}, {Frailis}, {Franceschi}, {Galeotta}, {Ganga}, {Giard}, {Giardino},
  {Giraud-H{\'e}raud}, {Gonz{\'a}lez-Nuevo}, {G{\'o}rski}, {Gratton},
  {Gregorio}, {Gruppuso}, {Hansen}, {Hanson}, {Harrison},
  {Henrot-Versill{\'e}}, {Hern{\'a}ndez-Monteagudo}, {Herranz}, {Hildebrandt},
  {Hivon}, {Hobson}, {Holmes}, {Hornstrup}, {Hovest}, {Huffenberger}, {Jaffe},
  {Jaffe}, {Jones}, {Juvela}, {Keih{\"a}nen}, {Keskitalo}, {Kisner}, {Kneissl},
  {Knoche}, {Knox}, {Kunz}, {Kurki-Suonio}, {Lagache}, {L{\"a}hteenm{\"a}ki},
  {Lamarre}, {Lasenby}, {Laureijs}, {Lawrence}, {Leahy}, {Leonardi},
  {Le{\'o}n-Tavares}, {Leroy}, {Lesgourgues}, {Liguori}, {Lilje},
  {Linden-V{\o}rnle}, {L{\'o}pez-Caniego}, {Lubin}, {Mac{\'\i}as-P{\'e}rez},
  {Maffei}, {Maino}, {Mandolesi}, {Maris}, {Marshall}, {Martin},
  {Mart{\'\i}nez-Gonz{\'a}lez}, {Masi}, {Massardi}, {Matarrese}, {Matthai},
  {Mazzotta}, {McGehee}, {Meinhold}, {Melchiorri}, {Mendes}, {Mennella},
  {Migliaccio}, {Mitra}, {Miville-Desch{\^e}nes}, {Moneti}, {Montier},
  {Morgante}, {Mortlock}, {Munshi}, {Murphy}, {Naselsky}, {Nati}, {Natoli},
  {Negrello}, {Netterfield}, {N{\o}rgaard-Nielsen}, {Noviello}, {Novikov},
  {Novikov}, {O'Dwyer}, {Osborne}, {Oxborrow}, {Paci}, {Pagano}, {Pajot},
  {Paladini}, {Paoletti}, {Partridge}, {Pasian}, {Patanchon}, {Pearson},
  {Perdereau}, {Perotto}, {Perrotta}, {Piacentini}, {Piat}, {Pierpaoli},
  {Pietrobon}, {Plaszczynski}, {Pointecouteau}, {Polenta}, {Ponthieu}, {Popa},
  {Poutanen}, {Pratt}, {Pr{\'e}zeau}, {Prunet}, {Puget}, {Rachen}, {Reach},
  {Rebolo}, {Reinecke}, {Remazeilles}, {Renault}, {Ricciardi}, {Riller},
  {Ristorcelli}, {Rocha}, {Rosset}, {Roudier}, {Rowan-Robinson},
  {Rubi{\~n}o-Mart{\'\i}n}, {Rusholme}, {Sandri}, {Santos}, {Savini}, {Scott},
  {Seiffert}, {Shellard}, {Spencer}, {Starck}, {Stolyarov}, {Stompor},
  {Sudiwala}, {Sunyaev}, {Sureau}, {Sutton}, {Suur-Uski}, {Sygnet}, {Tauber},
  {Tavagnacco}, {Terenzi}, {Toffolatti}, {Tomasi}, {Tristram}, {Tucci},
  {Tuovinen}, {T{\"u}rler}, {Umana}, {Valenziano}, {Valiviita}, {Van Tent},
  {Varis}, {Vielva}, {Villa}, {Vittorio}, {Wade}, {Walter}, {Wandelt}, {Yvon},
  {Zacchei}, \& {Zonca}}]{pccs}
{Planck Collaboration}, {Ade}, P.~A.~R., {Aghanim}, N., {et~al.} 2014, \aap,
  571, A28, \dodoi{10.1051/0004-6361/201321524}

\bibitem[{{Planck Collaboration} {et~al.}(2015){Planck Collaboration},
  {Aghanim}, {Altieri}, {Arnaud}, {Ashdown}, {Aumont}, {Baccigalupi}, {Banday},
  {Barreiro}, {Bartolo}, {Battaner}, {Beelen}, {Benabed}, {Benoit-L{\'e}vy},
  {Bernard}, {Bersanelli}, {Bethermin}, {Bielewicz}, {Bonavera}, {Bond},
  {Borrill}, {Bouchet}, {Boulanger}, {Burigana}, {Calabrese}, {Canameras},
  {Cardoso}, {Catalano}, {Chamballu}, {Chary}, {Chiang}, {Christensen},
  {Clements}, {Colombi}, {Couchot}, {Crill}, {Curto}, {Danese}, {Dassas},
  {Davies}, {Davis}, {de Bernardis}, {de Rosa}, {de Zotti}, {Delabrouille},
  {Diego}, {Dole}, {Donzelli}, {Dor{\'e}}, {Douspis}, {Ducout}, {Dupac},
  {Efstathiou}, {Elsner}, {En{\ss}lin}, {Falgarone}, {Flores-Cacho}, {Forni},
  {Frailis}, {Fraisse}, {Franceschi}, {Frejsel}, {Frye}, {Galeotta}, {Galli},
  {Ganga}, {Giard}, {Gjerl{\o}w}, {Gonz{\'a}lez-Nuevo}, {G{\'o}rski},
  {Gregorio}, {Gruppuso}, {Gu{\'e}ry}, {Hansen}, {Hanson}, {Harrison}, {Helou},
  {Hern{\'a}ndez-Monteagudo}, {Hildebrandt}, {Hivon}, {Hobson}, {Holmes},
  {Hovest}, {Huffenberger}, {Hurier}, {Jaffe}, {Jaffe}, {Keih{\"a}nen},
  {Keskitalo}, {Kisner}, {Kneissl}, {Knoche}, {Kunz}, {Kurki-Suonio},
  {Lagache}, {Lamarre}, {Lasenby}, {Lattanzi}, {Lawrence}, {Le Floc'h},
  {Leonardi}, {Levrier}, {Liguori}, {Lilje}, {Linden-V{\o}rnle},
  {L{\'o}pez-Caniego}, {Lubin}, {Mac{\'\i}as-P{\'e}rez}, {MacKenzie}, {Maffei},
  {Mandolesi}, {Maris}, {Martin}, {Martinache}, {Mart{\'\i}nez-Gonz{\'a}lez},
  {Masi}, {Matarrese}, {Mazzotta}, {Melchiorri}, {Mennella}, {Migliaccio},
  {Moneti}, {Montier}, {Morgante}, {Mortlock}, {Munshi}, {Murphy}, {Natoli},
  {Negrello}, {Nesvadba}, {Novikov}, {Novikov}, {Omont}, {Pagano}, {Pajot},
  {Pasian}, {Perdereau}, {Perotto}, {Perrotta}, {Pettorino}, {Piacentini},
  {Piat}, {Plaszczynski}, {Pointecouteau}, {Polenta}, {Popa}, {Pratt},
  {Prunet}, {Puget}, {Rachen}, {Reach}, {Reinecke}, {Remazeilles}, {Renault},
  {Ristorcelli}, {Rocha}, {Roudier}, {Rusholme}, {Sandri}, {Santos}, {Savini},
  {Scott}, {Spencer}, {Stolyarov}, {Sunyaev}, {Sutton}, {Sygnet}, {Tauber},
  {Terenzi}, {Toffolatti}, {Tomasi}, {Tristram}, {Tucci}, {Umana},
  {Valenziano}, {Valiviita}, {Valtchanov}, {Van Tent}, {Vieira}, {Vielva},
  {Wade}, {Wandelt}, {Wehus}, {Welikala}, {Zacchei}, \&
  {Zonca}}]{planck_cold_source}
{Planck Collaboration}, {Aghanim}, N., {Altieri}, B., {et~al.} 2015, \aap, 582,
  A30, \dodoi{10.1051/0004-6361/201424790}

\bibitem[{{Planck Collaboration} {et~al.}(2016){Planck Collaboration}, {Ade},
  {Aghanim}, {Arnaud}, {Aumont}, {Baccigalupi}, {Banday}, {Barreiro},
  {Bartolo}, {Battaner}, {Benabed}, {Benoit-L{\'e}vy}, {Bernard}, {Bersanelli},
  {Bielewicz}, {Bonaldi}, {Bonavera}, {Bond}, {Borrill}, {Bouchet},
  {Boulanger}, {Burigana}, {Butler}, {Calabrese}, {Catalano}, {Chiang},
  {Christensen}, {Clements}, {Colombo}, {Couchot}, {Coulais}, {Crill}, {Curto},
  {Cuttaia}, {Danese}, {Davies}, {Davis}, {de Bernardis}, {de Rosa}, {de
  Zotti}, {Delabrouille}, {Dickinson}, {Diego}, {Dole}, {Dor{\'e}}, {Douspis},
  {Ducout}, {Dupac}, {Elsner}, {En{\ss}lin}, {Eriksen}, {Falgarone}, {Finelli},
  {Flores-Cacho}, {Frailis}, {Fraisse}, {Franceschi}, {Galeotta}, {Galli},
  {Ganga}, {Giard}, {Giraud-H{\'e}raud}, {Gjerl{\o}w}, {Gonz{\'a}lez-Nuevo},
  {G{\'o}rski}, {Gregorio}, {Gruppuso}, {Gudmundsson}, {Hansen}, {Harrison},
  {Helou}, {Hern{\'a}ndez-Monteagudo}, {Herranz}, {Hildebrandt}, {Hivon},
  {Hobson}, {Hornstrup}, {Hovest}, {Huffenberger}, {Hurier}, {Jaffe}, {Jaffe},
  {Keih{\"a}nen}, {Keskitalo}, {Kisner}, {Kneissl}, {Knoche}, {Kunz},
  {Kurki-Suonio}, {Lagache}, {Lamarre}, {Lasenby}, {Lattanzi}, {Lawrence},
  {Leonardi}, {Levrier}, {Liguori}, {Lilje}, {Linden-V{\o}rnle},
  {L{\'o}pez-Caniego}, {Lubin}, {Mac{\'\i}as-P{\'e}rez}, {Maffei}, {Maggio},
  {Maino}, {Mandolesi}, {Mangilli}, {Maris}, {Martin},
  {Mart{\'\i}nez-Gonz{\'a}lez}, {Masi}, {Matarrese}, {Melchiorri}, {Mennella},
  {Migliaccio}, {Mitra}, {Miville-Desch{\^e}nes}, {Moneti}, {Montier},
  {Morgante}, {Mortlock}, {Munshi}, {Murphy}, {Nati}, {Natoli}, {Nesvadba},
  {Noviello}, {Novikov}, {Novikov}, {Oxborrow}, {Pagano}, {Pajot}, {Paoletti},
  {Partridge}, {Pasian}, {Pearson}, {Perdereau}, {Perotto}, {Pettorino},
  {Piacentini}, {Piat}, {Plaszczynski}, {Pointecouteau}, {Polenta}, {Pratt},
  {Prunet}, {Puget}, {Rachen}, {Reinecke}, {Remazeilles}, {Renault}, {Renzi},
  {Ristorcelli}, {Rocha}, {Rosset}, {Rossetti}, {Roudier},
  {Rubi{\~n}o-Mart{\'\i}n}, {Rusholme}, {Sandri}, {Santos}, {Savelainen},
  {Savini}, {Scott}, {Spencer}, {Stolyarov}, {Stompor}, {Sudiwala}, {Sunyaev},
  {Suur-Uski}, {Sygnet}, {Tauber}, {Terenzi}, {Toffolatti}, {Tomasi},
  {Tristram}, {Tucci}, {T{\"u}rler}, {Umana}, {Valenziano}, {Valiviita}, {Van
  Tent}, {Vielva}, {Villa}, {Wade}, {Wandelt}, {Wehus}, {Welikala}, {Yvon},
  {Zacchei}, \& {Zonca}}]{planck_cold_source_v2}
{Planck Collaboration}, {Ade}, P.~A.~R., {Aghanim}, N., {et~al.} 2016, \aap,
  596, A100, \dodoi{10.1051/0004-6361/201527206}

\bibitem[{{Prescott} {et~al.}(2008){Prescott}, {Kashikawa}, {Dey}, \&
  {Matsuda}}]{prescott08}
{Prescott}, M. K.~M., {Kashikawa}, N., {Dey}, A., \& {Matsuda}, Y. 2008, \apjl,
  678, L77, \dodoi{10.1086/588606}

\bibitem[{{Sanders} {et~al.}(2007){Sanders}, {Salvato}, {Aussel}, {Ilbert},
  {Scoville}, {Surace}, {Frayer}, {Sheth}, {Helou}, {Brooke}, {Bhattacharya},
  {Yan}, {Kartaltepe}, {Barnes}, {Blain}, {Calzetti}, {Capak}, {Carilli},
  {Carollo}, {Comastri}, {Daddi}, {Ellis}, {Elvis}, {Fall}, {Franceschini},
  {Giavalisco}, {Hasinger}, {Impey}, {Koekemoer}, {Le F{\`e}vre}, {Lilly},
  {Liu}, {McCracken}, {Mobasher}, {Renzini}, {Rich}, {Schinnerer}, {Shopbell},
  {Taniguchi}, {Thompson}, {Urry}, \& {Williams}}]{sanders07}
{Sanders}, D.~B., {Salvato}, M., {Aussel}, H., {et~al.} 2007, \apjs, 172, 86,
  \dodoi{10.1086/517885}

\bibitem[{{Sawicki} {et~al.}(2019){Sawicki}, {Arnouts}, {Huang}, {Coupon},
  {Golob}, {Gwyn}, {Foucaud}, {Moutard}, {Iwata}, {Liu}, {Chen}, {Desprez},
  {Harikane}, {Ono}, {Strauss}, {Tanaka}, {Thibert}, {Balogh}, {Bundy},
  {Chapman}, {Gunn}, {Hsieh}, {Ilbert}, {Jing}, {LeF{\`e}vre}, {Li}, {Matsuda},
  {Miyazaki}, {Nagao}, {Nishizawa}, {Ouchi}, {Shimasaku}, {Silverman}, {de la
  Torre}, {Tresse}, {Wang}, {Willott}, {Yamada}, {Yang}, \& {Yee}}]{sawicki19}
{Sawicki}, M., {Arnouts}, S., {Huang}, J., {et~al.} 2019, \mnras, 489, 5202,
  \dodoi{10.1093/mnras/stz2522}

\bibitem[{{Schinnerer} {et~al.}(2010){Schinnerer}, {Sargent}, {Bondi},
  {Smol{\v{c}}i{\'c}}, {Datta}, {Carilli}, {Bertoldi}, {Blain}, {Ciliegi},
  {Koekemoer}, \& {Scoville}}]{schinnerer10}
{Schinnerer}, E., {Sargent}, M.~T., {Bondi}, M., {et~al.} 2010, \apjs, 188,
  384, \dodoi{10.1088/0067-0049/188/2/384}

\bibitem[{{Shah} {et~al.}(2020){Shah}, {Kartaltepe}, {Magagnoli}, {Cox},
  {Wetherell}, {Vanderhoof}, {Calabro}, {Chartab}, {Conselice}, {Croton},
  {Donley}, {de Groot}, {de la Vega}, {Hathi}, {Ilbert}, {Inami}, {Kocevski},
  {Koekemoer}, {Lemaux}, {Mantha}, {Marchesi}, {Martig}, {Masters}, {McGrath},
  {McIntosh}, {Moreno}, {Nayyeri}, {Pampliega}, {Salvato}, {Snyder},
  {Straughn}, {Treister}, \& {Weston}}]{shah20}
{Shah}, E.~A., {Kartaltepe}, J.~S., {Magagnoli}, C.~T., {et~al.} 2020, \apj,
  904, 107, \dodoi{10.3847/1538-4357/abbf59}

\bibitem[{{Shen} {et~al.}(2021){Shen}, {Lemaux}, {Lubin}, {Cucciati}, {Le
  F{\`e}vre}, {Liu}, {Fang}, {Pelliccia}, {Tomczak}, {McKean}, {Miller},
  {Fassnacht}, {Gal}, {Hung}, {Hathi}, {Bardelli}, {Vergani}, \&
  {Zucca}}]{shen21}
{Shen}, L., {Lemaux}, B.~C., {Lubin}, L.~M., {et~al.} 2021, \apj, 912, 60,
  \dodoi{10.3847/1538-4357/abee75}

\bibitem[{{Shi} {et~al.}(2019{\natexlab{a}}){Shi}, {Huang}, {Lee}, {Toshikawa},
  {Bowen}, {Malavasi}, {Lemaux}, {Cucciati}, {Le Fevre}, \& {Dey}}]{shi19}
{Shi}, K., {Huang}, Y., {Lee}, K.-S., {et~al.} 2019{\natexlab{a}}, \apj, 879,
  9, \dodoi{10.3847/1538-4357/ab2118}

\bibitem[{{Shi} {et~al.}(2019{\natexlab{b}}){Shi}, {Lee}, {Dey}, {Huang},
  {Malavasi}, {Hung}, {Inami}, {Ashby}, {Duncan}, {Xue}, {Reddy}, {Hong},
  {Jannuzi}, {Cooper}, {Gonzalez}, {R{\"o}ttgering}, {Best}, \&
  {Tasse}}]{shi19a}
{Shi}, K., {Lee}, K.-S., {Dey}, A., {et~al.} 2019{\natexlab{b}}, \apj, 871, 83,
  \dodoi{10.3847/1538-4357/aaf85d}

\bibitem[{{Shimakawa} {et~al.}(2014){Shimakawa}, {Kodama}, {Tadaki}, {Tanaka},
  {Hayashi}, \& {Koyama}}]{shimakawa14}
{Shimakawa}, R., {Kodama}, T., {Tadaki}, K.~I., {et~al.} 2014, \mnras, 441, L1,
  \dodoi{10.1093/mnrasl/slu029}

\bibitem[{{Shimakawa} {et~al.}(2018){Shimakawa}, {Kodama}, {Hayashi},
  {Prochaska}, {Tanaka}, {Cai}, {Suzuki}, {Tadaki}, \& {Koyama}}]{shimakawa18}
{Shimakawa}, R., {Kodama}, T., {Hayashi}, M., {et~al.} 2018, \mnras, 473, 1977,
  \dodoi{10.1093/mnras/stx2494}

\bibitem[{{Snyder} {et~al.}(2012){Snyder}, {Brodwin}, {Mancone}, {Zeimann},
  {Stanford}, {Gonzalez}, {Stern}, {Eisenhardt}, {Brown}, {Dey}, {Jannuzi}, \&
  {Perlmutter}}]{snyder12}
{Snyder}, G.~F., {Brodwin}, M., {Mancone}, C.~M., {et~al.} 2012, \apj, 756,
  114, \dodoi{10.1088/0004-637X/756/2/114}

\bibitem[{{Somerville} \& {Dav{\'e}}(2015)}]{somerville15}
{Somerville}, R.~S., \& {Dav{\'e}}, R. 2015, \araa, 53, 51,
  \dodoi{10.1146/annurev-astro-082812-140951}

\bibitem[{{Stanford} {et~al.}(1998){Stanford}, {Eisenhardt}, \&
  {Dickinson}}]{stanford98}
{Stanford}, S.~A., {Eisenhardt}, P.~R., \& {Dickinson}, M. 1998, \apj, 492,
  461, \dodoi{10.1086/305050}

\bibitem[{{Steidel} {et~al.}(1998){Steidel}, {Adelberger}, {Dickinson},
  {Giavalisco}, {Pettini}, \& {Kellogg}}]{steidel98}
{Steidel}, C.~C., {Adelberger}, K.~L., {Dickinson}, M., {et~al.} 1998, \apj,
  492, 428, \dodoi{10.1086/305073}

\bibitem[{{Stevens} {et~al.}(2010){Stevens}, {Jarvis}, {Coppin}, {Page},
  {Greve}, {Carrera}, \& {Ivison}}]{stevens10}
{Stevens}, J.~A., {Jarvis}, M.~J., {Coppin}, K.~E.~K., {et~al.} 2010, \mnras,
  405, 2623, \dodoi{10.1111/j.1365-2966.2010.16641.x}

\bibitem[{{Topping} {et~al.}(2018){Topping}, {Shapley}, {Steidel}, {Naoz}, \&
  {Primack}}]{topping18}
{Topping}, M.~W., {Shapley}, A.~E., {Steidel}, C.~C., {Naoz}, S., \& {Primack},
  J.~R. 2018, \apj, 852, 134, \dodoi{10.3847/1538-4357/aa9f0f}

\bibitem[{{Toshikawa} {et~al.}(2012){Toshikawa}, {Kashikawa}, {Ota},
  {Morokuma}, {Shibuya}, {Hayashi}, {Nagao}, {Jiang}, {Malkan}, {Egami},
  {Shimasaku}, {Motohara}, \& {Ishizaki}}]{toshikawa12}
{Toshikawa}, J., {Kashikawa}, N., {Ota}, K., {et~al.} 2012, \apj, 750, 137,
  \dodoi{10.1088/0004-637X/750/2/137}

\bibitem[{{Toshikawa} {et~al.}(2016){Toshikawa}, {Kashikawa}, {Overzier},
  {Malkan}, {Furusawa}, {Ishikawa}, {Onoue}, {Ota}, {Tanaka}, {Niino}, \&
  {Uchiyama}}]{toshikawa16}
{Toshikawa}, J., {Kashikawa}, N., {Overzier}, R., {et~al.} 2016, \apj, 826,
  114, \dodoi{10.3847/0004-637X/826/2/114}

\bibitem[{{Toshikawa} {et~al.}(2018){Toshikawa}, {Uchiyama}, {Kashikawa},
  {Ouchi}, {Overzier}, {Ono}, {Harikane}, {Ishikawa}, {Kodama}, {Matsuda},
  {Lin}, {Onoue}, {Tanaka}, {Nagao}, {Akiyama}, {Komiyama}, {Goto}, \&
  {Lee}}]{toshikawa18}
{Toshikawa}, J., {Uchiyama}, H., {Kashikawa}, N., {et~al.} 2018, \pasj, 70,
  S12, \dodoi{10.1093/pasj/psx102}

\bibitem[{{Tozzi} {et~al.}(2022){Tozzi}, {Pentericci}, {Gilli}, {Pannella},
  {Fiore}, {Miley}, {Nonino}, {Rottgering}, {Strazzullo}, {Anderson},
  {Borgani}, {Calabro'}, {Carilli}, {Dannerbauer}, {Di Mascolo}, {Feruglio},
  {Gobat}, {Jin}, {Liu}, {Mroczkowski}, {Norman}, {Rasia}, {Rosati}, \&
  {Saro}}]{tozzi22}
{Tozzi}, P., {Pentericci}, L., {Gilli}, R., {et~al.} 2022, arXiv e-prints,
  arXiv:2203.02208.
\newblock \doarXiv{2203.02208}

\bibitem[{{Trainor} \& {Steidel}(2012)}]{trainor12}
{Trainor}, R.~F., \& {Steidel}, C.~C. 2012, \apj, 752, 39,
  \dodoi{10.1088/0004-637X/752/1/39}

\bibitem[{{Umehata} {et~al.}(2019){Umehata}, {Fumagalli}, {Smail}, {Matsuda},
  {Swinbank}, {Cantalupo}, {Sykes}, {Ivison}, {Steidel}, {Shapley}, {Vernet},
  {Yamada}, {Tamura}, {Kubo}, {Nakanishi}, {Kajisawa}, {Hatsukade}, \&
  {Kohno}}]{umehata19}
{Umehata}, H., {Fumagalli}, M., {Smail}, I., {et~al.} 2019, Science, 366, 97,
  \dodoi{10.1126/science.aaw5949}

\bibitem[{{Vanden Berk} {et~al.}(2001){Vanden Berk}, {Richards}, {Bauer},
  {Strauss}, {Schneider}, {Heckman}, {York}, {Hall}, {Fan}, {Knapp},
  {Anderson}, {Annis}, {Bahcall}, {Bernardi}, {Briggs}, {Brinkmann}, {Brunner},
  {Burles}, {Carey}, {Castander}, {Connolly}, {Crocker}, {Csabai}, {Doi},
  {Finkbeiner}, {Friedman}, {Frieman}, {Fukugita}, {Gunn}, {Hennessy},
  {Ivezi{\'c}}, {Kent}, {Kunszt}, {Lamb}, {Leger}, {Long}, {Loveday}, {Lupton},
  {Meiksin}, {Merelli}, {Munn}, {Newberg}, {Newcomb}, {Nichol}, {Owen}, {Pier},
  {Pope}, {Rockosi}, {Schlegel}, {Siegmund}, {Smee}, {Snir}, {Stoughton},
  {Stubbs}, {SubbaRao}, {Szalay}, {Szokoly}, {Tremonti}, {Uomoto}, {Waddell},
  {Yanny}, \& {Zheng}}]{vandenberk01}
{Vanden Berk}, D.~E., {Richards}, G.~T., {Bauer}, A., {et~al.} 2001, \aj, 122,
  549, \dodoi{10.1086/321167}

\bibitem[{{Venemans} {et~al.}(2002){Venemans}, {Kurk}, {Miley},
  {R{\"o}ttgering}, {van Breugel}, {Carilli}, {De Breuck}, {Ford}, {Heckman},
  {McCarthy}, \& {Pentericci}}]{venemans02}
{Venemans}, B.~P., {Kurk}, J.~D., {Miley}, G.~K., {et~al.} 2002, \apjl, 569,
  L11, \dodoi{10.1086/340563}

\bibitem[{{Venemans} {et~al.}(2007){Venemans}, {R{\"o}ttgering}, {Miley}, {van
  Breugel}, {de Breuck}, {Kurk}, {Pentericci}, {Stanford}, {Overzier}, {Croft},
  \& {Ford}}]{venemans07}
{Venemans}, B.~P., {R{\"o}ttgering}, H.~J.~A., {Miley}, G.~K., {et~al.} 2007,
  \aap, 461, 823, \dodoi{10.1051/0004-6361:20053941}

\bibitem[{{Volonteri} {et~al.}(2003){Volonteri}, {Haardt}, \&
  {Madau}}]{volonteri03}
{Volonteri}, M., {Haardt}, F., \& {Madau}, P. 2003, \apj, 582, 559,
  \dodoi{10.1086/344675}

\bibitem[{{von der Linden} {et~al.}(2010){von der Linden}, {Wild}, {Kauffmann},
  {White}, \& {Weinmann}}]{vonderlinden10}
{von der Linden}, A., {Wild}, V., {Kauffmann}, G., {White}, S. D.~M., \&
  {Weinmann}, S. 2010, \mnras, 404, 1231,
  \dodoi{10.1111/j.1365-2966.2010.16375.x}

\bibitem[{{Wang} {et~al.}(2016){Wang}, {Elbaz}, {Daddi}, {Finoguenov}, {Liu},
  {Schreiber}, {Mart{\'{\i}}n}, {Strazzullo}, {Valentino}, {van der Burg},
  {Zanella}, {Ciesla}, {Gobat}, {Le Brun}, {Pannella}, {Sargent}, {Shu}, {Tan},
  {Cappelluti}, \& {Li}}]{wang16}
{Wang}, T., {Elbaz}, D., {Daddi}, E., {et~al.} 2016, \apj, 828, 56,
  \dodoi{10.3847/0004-637X/828/1/56}

\bibitem[{{Weaver} {et~al.}(2022){Weaver}, {Kauffmann}, {Ilbert}, {McCracken},
  {Moneti}, {Toft}, {Brammer}, {Shuntov}, {Davidzon}, {Hsieh}, {Laigle},
  {Anastasiou}, {Jespersen}, {Vinther}, {Capak}, {Casey}, {McPartland},
  {Milvang-Jensen}, {Mobasher}, {Sanders}, {Zalesky}, {Arnouts}, {Aussel},
  {Dunlop}, {Faisst}, {Franx}, {Furtak}, {Fynbo}, {Gould}, {Greve}, {Gwyn},
  {Kartaltepe}, {Kashino}, {Koekemoer}, {Kokorev}, {Le F{\`e}vre}, {Lilly},
  {Masters}, {Magdis}, {Mehta}, {Peng}, {Riechers}, {Salvato}, {Sawicki},
  {Scarlata}, {Scoville}, {Shirley}, {Silverman}, {Sneppen}, {Smolc̆i{\'c}},
  {Steinhardt}, {Stern}, {Tanaka}, {Taniguchi}, {Teplitz}, {Vaccari}, {Wang},
  \& {Zamorani}}]{weaver22}
{Weaver}, J.~R., {Kauffmann}, O.~B., {Ilbert}, O., {et~al.} 2022, \apjs, 258,
  11, \dodoi{10.3847/1538-4365/ac3078}

\bibitem[{{Wold} {et~al.}(2003){Wold}, {Armus}, {Neugebauer}, {Jarrett}, \&
  {Lehnert}}]{wold03}
{Wold}, M., {Armus}, L., {Neugebauer}, G., {Jarrett}, T.~H., \& {Lehnert},
  M.~D. 2003, \aj, 126, 1776, \dodoi{10.1086/378362}

\bibitem[{{Yang} {et~al.}(2010){Yang}, {Zabludoff}, {Eisenstein}, \&
  {Dav{\'e}}}]{yang10}
{Yang}, Y., {Zabludoff}, A., {Eisenstein}, D., \& {Dav{\'e}}, R. 2010, \apj,
  719, 1654, \dodoi{10.1088/0004-637X/719/2/1654}

\bibitem[{{Yang} {et~al.}(2009){Yang}, {Zabludoff}, {Tremonti}, {Eisenstein},
  \& {Dav{\'e}}}]{yang09}
{Yang}, Y., {Zabludoff}, A., {Tremonti}, C., {Eisenstein}, D., \& {Dav{\'e}},
  R. 2009, \apj, 693, 1579, \dodoi{10.1088/0004-637X/693/2/1579}

\bibitem[{{Zheng} {et~al.}(2021){Zheng}, {Cai}, {An}, {Fan}, \&
  {Shi}}]{zheng20}
{Zheng}, X.~Z., {Cai}, Z., {An}, F.~X., {Fan}, X., \& {Shi}, D.~D. 2021,
  \mnras, 500, 4354, \dodoi{10.1093/mnras/staa2882}

\end{thebibliography}

\end{document}